\documentclass[pdflatex,sn-mathphys-num]{sn-jnl}

\usepackage[T1]{fontenc}
\usepackage[utf8]{inputenc}

\usepackage{graphicx}%
\usepackage{multirow}%
\usepackage{amsmath,amssymb,amsfonts}%
\usepackage{amsthm}%
\usepackage{mathrsfs}%
\usepackage{xcolor}%
\usepackage{textcomp}%
\usepackage{colortbl}
\usepackage{array}
\usepackage{makecell}
\usepackage{manyfoot}%
\usepackage{booktabs}%
\usepackage{longtable}%
\usepackage{ragged2e}%
\usepackage{algorithm}%
\usepackage{algorithmicx}%
\usepackage{algpseudocode}%
\usepackage{listings}%
\usepackage{CJKutf8}
\usepackage{wasysym}
\usepackage[most]{tcolorbox}
\usepackage{booktabs}
\usepackage{multirow}

\definecolor{queryboxbg}{RGB}{247,247,247}
\definecolor{queryboxframe}{RGB}{210,210,210}
\definecolor{caseboxbg}{RGB}{242,248,255}
\definecolor{caseboxframe}{RGB}{120,160,205}
\definecolor{evolveboxbg}{RGB}{242,250,244}
\definecolor{evolveboxframe}{RGB}{112,166,128}
\definecolor{promptboxbg}{RGB}{247,247,247}
\definecolor{promptboxframe}{RGB}{80,80,80}
\newtcolorbox{querybox}{
  enhanced,
  colback=queryboxbg,
  colframe=queryboxframe,
  boxrule=0.5pt,
  arc=4pt,
  left=7pt,
  right=7pt,
  top=6pt,
  bottom=6pt,
  before skip=7pt,
  after skip=7pt
}
\newtcolorbox{casebox}[1]{
  enhanced,
  breakable,
  fontupper=\RaggedRight,
  colback=caseboxbg,
  colframe=caseboxframe,
  colbacktitle=caseboxframe,
  coltitle=white,
  fonttitle=\bfseries,
  title={#1},
  boxrule=0.7pt,
  arc=3pt,
  left=7pt,
  right=7pt,
  top=6pt,
  bottom=6pt,
  before skip=12pt,
  after skip=14pt
}
\newtcolorbox{evolvebox}[1]{
  enhanced,
  breakable,
  fontupper=\RaggedRight,
  colback=evolveboxbg,
  colframe=evolveboxframe,
  colbacktitle=evolveboxframe,
  coltitle=white,
  fonttitle=\bfseries,
  title={#1},
  boxrule=0.7pt,
  arc=3pt,
  left=7pt,
  right=7pt,
  top=6pt,
  bottom=6pt,
  before skip=12pt,
  after skip=14pt
}
\newtcolorbox{promptbox}[1]{
  enhanced,
  breakable,
  fontupper=\RaggedRight,
  colback=promptboxbg,
  colframe=promptboxframe,
  colbacktitle=promptboxframe,
  coltitle=white,
  fonttitle=\bfseries,
  title={#1},
  boxrule=0.7pt,
  arc=3pt,
  left=7pt,
  right=7pt,
  top=6pt,
  bottom=6pt,
  before skip=12pt,
  after skip=14pt
}

\theoremstyle{thmstyleone}%

\theoremstyle{thmstyletwo}%

\theoremstyle{thmstylethree}%

\begin{document}

\title[PaSaMaster]{Towards Recursive Self-Evolving Agentic Literature Retrieval}

\author*[1,2]{\fnm{Yuwen} \sur{Du}}\email{sjtu0267098@sjtu.edu.cn}
\equalcont{These authors contributed equally to this work.}

\author[1,2]{\fnm{Tian} \sur{Jin}}
\equalcont{These authors contributed equally to this work.}

\author[1,2]{\fnm{Jing} \sur{Kang}}

\author[1,2]{\fnm{Xianghe} \sur{Pang}}

\author[1,2]{\fnm{Jingyi} \sur{Chai}}

\author[1,2]{\fnm{Tingjia} \sur{Miao}}

\author[1,2]{\fnm{Fenyi} \sur{Liu}}

\author[3]{\fnm{WenHao} \sur{Wang}}

\author[2]{\fnm{Sikai} \sur{Yao}}

\author[2]{\fnm{Yuzhi} \sur{Zhang}}

\author*[1,2]{\fnm{Siheng} \sur{Chen}}\email{sihengc@sjtu.edu.cn}

\affil[1]{\orgname{Shanghai Jiao Tong University}, \orgaddress{\city{Shanghai}, \country{China}}}

\affil[2]{\orgname{SciLand}, \orgaddress{\city{Shanghai}, \country{China}}}

\affil[3]{\orgname{Zhejiang University}, \orgaddress{\city{Hangzhou}, \country{China}}}

\abstract{Scientific literature retrieval must understand complex search intents while preserving source authenticity. Traditional keyword and embedding-based systems return authentic sources but miss nuanced intents, whereas large language models capture richer intents but may fabricate citations. We introduce PaSaMaster, a Recursive Self-Evolving agentic literature retrieval system that iteratively analyzes intent, retrieves verified papers and ranks them with evidence-grounded relevance scores. PaSaMaster combines self-evolving retrieval that refines search intent from ranked evidence over time, hallucination-free ranking over verified papers rather than generated citations, and cost-efficient planning--retrieval separation that reserves frontier LLMs for intent understanding while delegating retrieval and scoring to lightweight models and customized corpora. Across 38 disciplines in PaSaMaster-Bench, PaSaMaster achieves a 16.5$\times$ higher F1-score than Google Scholar and a 37.8\% higher F1-score than GPT-5.2 at about 1\% of the cost, while reducing source hallucination from 32.66\% in generative LLMs to zero: \url{https://github.com/sjtu-sai-agents/PaSaMaster}}


\maketitle

\begin{table*}[!tbp]
\centering
\small
\setlength{\tabcolsep}{5pt}
\renewcommand{\arraystretch}{1.8}
\newcolumntype{L}[1]{>{\raggedright\arraybackslash}p{#1}}
\caption{\textbf{The Five Paradigms of Scientific Literature Discovery.} 
Existing paradigms each improve one aspect of literature retrieval, but fail to jointly achieve adaptive intent understanding, hallucination-free evidence grounding, and cost-efficient scaling. PaSaMaster resolves these limitations through agentic Recursive Self-Evolving, evidence-grounded retrieval.}
\label{tab:five_level}
\scalebox{0.6}{
\begin{tabular}{|L{4.2cm}|L{4.6cm}|L{3.6cm}|L{3.6cm}|L{3.4cm}|}
\hline
\rowcolor{gray!18}
\textbf{Paradigm Level} 
& \textbf{Representative Systems} 
& \textbf{Intent Adaptivity} 
& \textbf{Source Reliability} 
& \textbf{Cost Efficiency} \\
\hline

\textbf{Level 0: Lexical Retrieval}
& Google Scholar~\cite{google_scholar}, PubMed~\cite{pubmed-ncbi-official}
& Keyword-based; severe intent compression 
& Verified indexed papers 
& Efficient but semantically shallow \\
\hline

\textbf{Level 1: Semantic Retrieval}
& OpenScholar~\cite{asai2024openscholar}, Bohrium Navigator~\cite{zhang2025bohriumscimaster}
& Passive embedding matching; limited intent compression
& Verified indexed papers 
& Efficient but semantically shallow \\
\hline

\textbf{Level 2: Generative LLMs}
& GPT-5.2~\cite{openai2026gpt54}, Gemini 3.1 Pro~\cite{google2026gemini31pro}, DeepSeek~\cite{deepseekai2025deepseekv32}
& Strong natural-language understanding 
& Prone to hallucinated papers 
& Expensive to deploy at scale \\
\hline

\textbf{Level 3: Fixed-Pipeline Agentic Retrieval}
& Google Scholar Labs~\cite{google2025scholarlabs}, PaSa~\cite{he2025pasa}
& User intent fixed at the outset; no cognition update during retrieval
& Verified indexed papers 
& Cost-controlled, but constrained by fixed intent interpretation \\
\hline

\rowcolor{gray!10}
\textbf{Level 4: Recursive Self-Evolving Agentic Retrieval}
& \textbf{PaSaMaster (Ours)}
& \textbf{Iteratively refines intent using ranked evidence} 
& \textbf{Verified indexed papers} 
& \textbf{Cost-efficient planning--retrieval separation} \\
\hline
\end{tabular}
}
\end{table*}

\begin{figure*}[!tbp]
    \centering
    \begin{minipage}{\linewidth}
        \textbf{a}\par\vspace{0.2em}
        \includegraphics[width=1\linewidth]{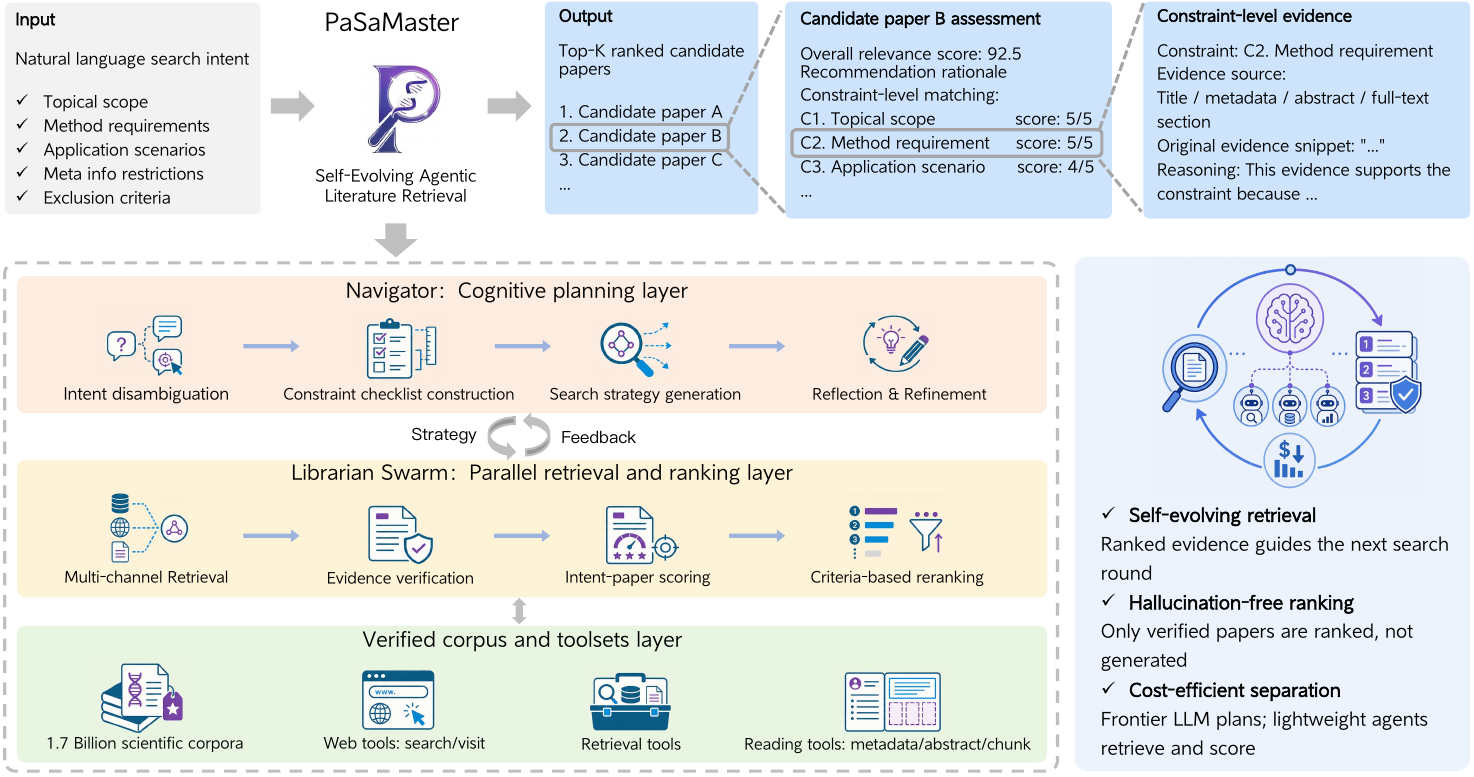}
    \end{minipage}
    \vspace{0.4em}
    \begin{minipage}{\linewidth}
        \textbf{b}\par\vspace{0.2em}
        \includegraphics[width=1\linewidth]{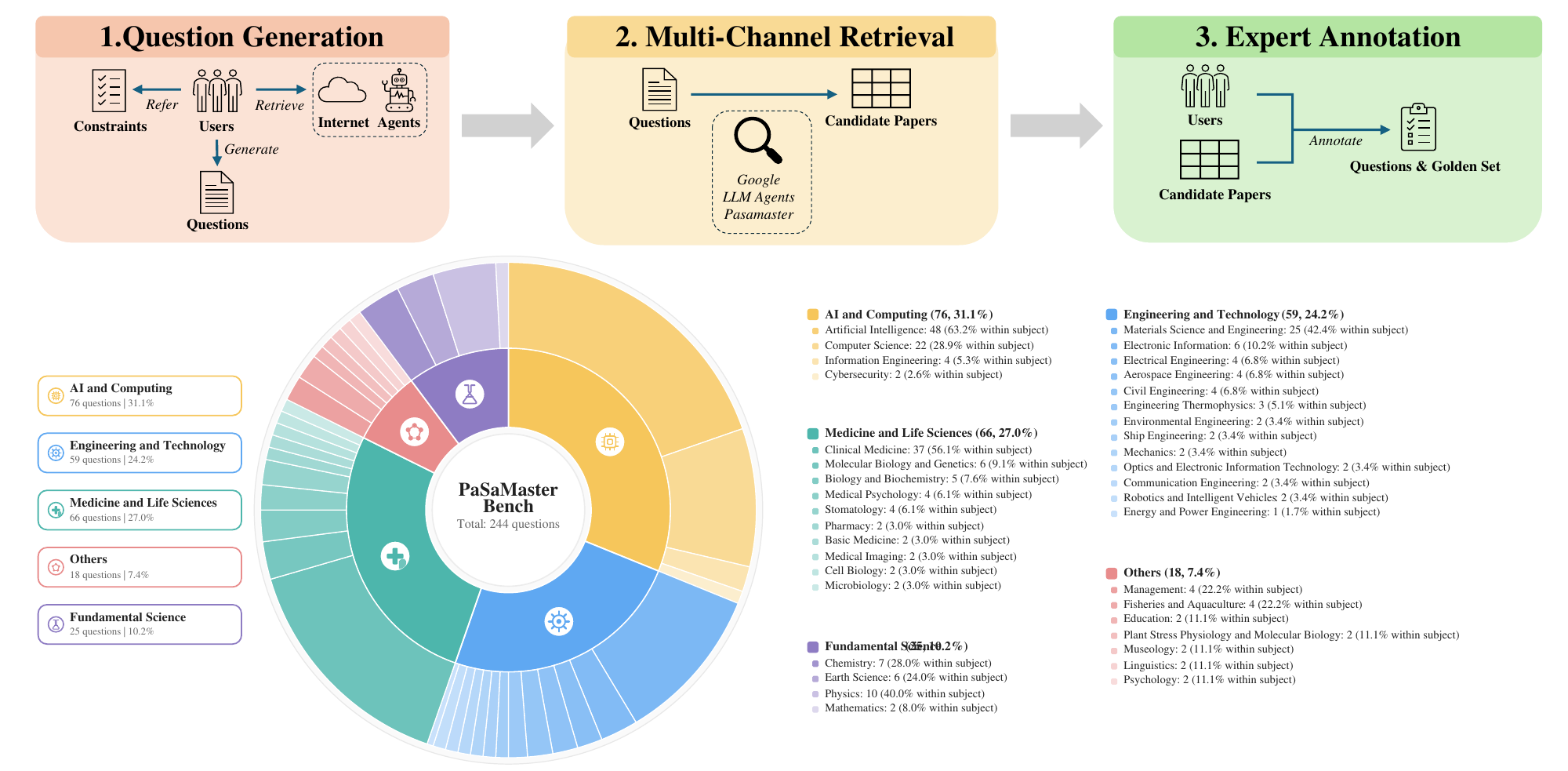}
    \end{minipage}
    \caption{\small \textbf{Overview of PaSaMaster and PaSaMaster-Bench.} \textbf{a,} PaSaMaster converts a complex natural-language search intent into a verified, evidence-ranked paper list through three layers: a Navigator that refines intent and search strategy, a Librarian Swarm that retrieves, verifies and scores candidate papers, and verified corpora and tools that ground all outputs in real sources. \textbf{b,} PaSaMaster-Bench evaluates complex literature retrieval through expert-written search intents, multi-channel candidate retrieval and checklist-based expert annotation. The benchmark contains 244 tasks across 38 disciplines, enabling evaluation of intent understanding, source reliability and ranking quality.}
    \label{fig:teaser}
\end{figure*}

Scientific literature retrieval is the axiomatic starting point of all scientific inquiry~\cite{gusenbauer2021searching}. Before formulating hypotheses, designing experiments, or building new theories, researchers must fundamentally navigate the vast and ever-expanding corpus of existing knowledge~\cite{fortunato2018science}. However, the volume of scientific publications has grown exponentially over recent decades, decisively overwhelming the fixed cognitive bandwidth of individual researchers~\cite{bornmann2015growth,Landhuis_2016,houssard2025gerontocratization}. This severe information overload has driven an inevitable reliance on artificial intelligence to automate and accelerate knowledge discovery~\cite{wang2023scientific}. More importantly, modern literature search is rarely a simple keyword lookup~\cite{furnas1987vocabulary,beel2010aseo}. Researchers often express complex academic intents involving technical constraints, application contexts, and implicit background knowledge~\cite{white2009exploratory,gusenbauer2021searching,ajith2024litsearch}. As large language models reshape scientific research workflows, literature retrieval therefore faces a new central challenge: \textit{how to deeply understand complex search intents while ensuring that every returned source is real and verifiable}~\cite{zhang2025llmscimethod,ajith2024litsearch}.

Existing literature retrieval systems still struggle to jointly achieve complex intent understanding and source authenticity. Some methods preserve source authenticity at the cost of shallow intent understanding~\cite{google_scholar,asai2024openscholar,zhang2025bohriumscimaster}, whereas others improve semantic comprehension while sacrificing factual reliability~\cite{deepseekai2025deepseekv32,kimiteam2026kimik2,minimax2025minimaxm1,glm5team2025glm45,google2026gemini31pro,openai2026gpt54}. This creates a persistent tradeoff between reliable but limited retrieval and more intelligent but less trustworthy literature discovery. 

The tradeoff between intent understanding and source reliability is clearer when viewed through the evolution of literature retrieval paradigms (Table~\ref{tab:five_level}). \textit{Level 0 (Lexical Retrieval)}~\cite{google_scholar, pubmed-ncbi-official} guarantees source authenticity through indexed databases, but reduces complex research intents to rigid keywords, causing severe intent compression. \textit{Level 1 (Semantic Retrieval)}~\cite{asai2024openscholar, zhang2025bohriumscimaster} improves over exact keyword matching by using embedding-based similarity and retrieval-augmented matching~\cite{karpukhin2020dense,lewis2021rag}, but still treats retrieval as passive query--document matching and lacks the ability to actively clarify, decompose, or refine complex intents. \textit{Level 2 (Generative LLMs)}~\cite{deepseekai2025deepseekv32, kimiteam2026kimik2, minimax2025minimaxm1, glm5team2025glm45, google2026gemini31pro, openai2026gpt54} offers stronger intent comprehension, yet its probabilistic generation introduces fabricated papers~\cite{zhang2025sirenssong,Farquhar_2024}, undermining the factual trust required for scientific inquiry. \textit{Level 3 (Fixed-Pipeline Agentic Retrieval)}~\cite{google2025scholarlabs,he2025pasa} mitigates source hallucination by grounding LLM agents in verifiable retrieval tools~\cite{nakano2021webgpt,schick2023toolformer,qin2023toolllm,yao2023react,du2026openseeker,du2026openseekerv2}. However, these systems typically follow a predefined retrieve--read--answer pipeline in which the interpretation of the user question is largely fixed at the beginning. For complex research intents, this initial interpretation can be incomplete or biased toward only part of the request, causing later retrieval steps to miss relevant subtopics or return papers that satisfy only some constraints. This limitation motivates retrieval systems that can update their understanding of the intent as evidence accumulates.

The unresolved need for adaptive, verifiable and efficient retrieval motivates \textit{Level 4 (Recursive Self-Evolving Agentic Retrieval)}, represented by \textbf{PaSaMaster}. PaSaMaster is a Recursive Self-Evolving agentic literature retrieval system that iteratively analyzes intent, retrieves verified papers and ranks them with evidence-grounded relevance scores. Rather than treating literature search as one-shot query--document matching problem, PaSaMaster formulates scientific literature discovery as a Recursive Self-Evolving intent--paper relevance ranking process. This design enables the system to align with complex research intents while ensuring that every returned source is real, verifiable, and grounded in customized corpora.

PaSaMaster is built on three key designs. First, Recursive Self-Evolving retrieval: it transforms literature retrieval from one-shot query--document matching into an adaptive search process that evolves over time~\cite{guo2024multiagent,shinn2023reflexion}, where retrieved and ranked evidence is used to identify coverage gaps, refine the research intent, and guide subsequent retrieval rounds. Second, hallucination-free ranking: it treats literature discovery as intent--paper relevance ranking rather than generation, and trains a lightweight Ranker on multidisciplinary query--paper evidence so that Librarian agents can make expert-style relevance judgments across disciplines while ranking only verified papers grounded in original evidence. Third, cost-efficient planning--retrieval separation: it uses frontier LLMs only for intent understanding and refinement, while delegating large-scale retrieval and relevance scoring to customized scientific corpora and lightweight models. Together, these designs enable PaSaMaster to align with complex research intents while maintaining source verifiability and cost-efficient scalability.

To evaluate retrieval capability on complex natural-language literature search problems, we introduce \textbf{PaSaMaster-Bench}, the first multidisciplinary literature retrieval benchmark designed for complex search intents. Unlike conventional retrieval benchmarks~\cite{he2025pasa,kang2025researcharenabenchmarkinglargelanguage,bragg2026astabenchrigorousbenchmarkingai,xiong2026autoresearchbenchbenchmarkingaiagents,ajith2024litsearch} built around short keyword queries, PaSaMaster-Bench focuses on highly specific, multi-constrained natural language search intents that require systems to search, verify, and rank all papers satisfying explicit criteria. The benchmark contains \textit{244 expert-curated tasks} spanning \textit{38 scientific disciplines}, with queries, constraints, target paper lists, and evaluation checklists annotated and verified by human domain experts.

The PaSaMaster-Bench evaluation reveals the severe inaccuracy and incompleteness of traditional keyword retrieval, with PaSaMaster improving F1-score by 16.5$\times$. The evaluation also exposes the unreliability of generative LLMs, which exhibit hallucination rates up to 32.66\%. Remarkably, PaSaMaster outperforms GPT-5.2 by 37.8\% while using only 1\% of its computational cost, and maintains zero source hallucination. These results demonstrate that Recursive Self-Evolving, evidence-grounded relevance ranking can improve complex intent understanding, eliminate source hallucination and enable low-cost scientific literature discovery at scale.

\section{Results}\label{sec2}

\subsection{PaSaMaster system overview}

Figure~\ref{fig:teaser}a summarizes how PaSaMaster converts a natural-language search intent into an auditable ranked paper list. The system first makes the implicit structure of the request explicit, turning topical scope, methodological requirements, application context, metadata restrictions and exclusion rules into a retrieval strategy and verification checklist. It then returns top-$K$ candidate papers with paper-level relevance scores, recommendation rationales and constraint-level judgments. Each judgment is linked to the evidence used to make it, including metadata, abstracts, full-text snippets and the reasoning connecting the evidence to the stated constraint.

PaSaMaster is organized into three layers. The Navigator performs intent disambiguation, checklist construction, search planning and round-by-round reflection. The Librarian Swarm executes multi-channel retrieval, evidence verification, intent--paper scoring and criteria-based reranking. The verified corpus and toolset layer supplies scientific corpora, search and visit tools, retrieval operators and reading tools for metadata, abstracts and evidence chunks. The strategy--feedback loop between the Navigator and Librarians makes retrieval self-evolving: ranked evidence from each round updates the system's understanding of the user's intent and determines the next search direction, while the verified corpus layer keeps all ranked papers authentic and evidence-grounded.

\subsection{PaSaMaster-Bench captures complex search intents}

PaSaMaster-Bench evaluates whether retrieval systems can resolve the kinds of compositional search intents that arise in real research work~\cite{white2009exploratory,gusenbauer2021searching,ajith2024litsearch}. Such intents are rarely reducible to a topic keyword: a user may simultaneously specify the scientific scope, required method, application domain, venue or time window, and exclusions that remove papers that look related but are not the intended target. Because these constraints are expressed in natural language rather than as explicit database filters, the system must infer the full intent before it can identify the correct paper set.

A representative task is: \textit{Could you help me find empirical studies on federated learning for multi-hospital collaborative training of medical imaging AI models, published in Nature Medicine, Nature Communications or IEEE Transactions on Medical Imaging between 2023 and October 2025?} This is the kind of request a researcher might make when preparing a review, grant proposal or related-work section. It requires the system to satisfy the topic, method, application, venue, time and exclusion constraints jointly; a paper that only discusses federated learning, or only studies medical imaging, is not sufficient.

The benchmark contains 244 independent literature discovery tasks across 38 scientific disciplines. For each task, domain experts write the natural-language intent, decompose it into objective checklist items, assemble a broad candidate pool through multiple retrieval channels and annotate every candidate against the checklist (Fig.~\ref{fig:teaser}b). The resulting target set is therefore stricter than topical relevance: a paper is correct only if it satisfies all required constraints. PaSaMaster-Bench measures whether a system can recover the intended paper set behind a realistic research question, not merely whether it can retrieve papers from the same field.

\subsection{Experimental setup}

We compare PaSaMaster with four groups of baselines. Lexical retrieval systems, represented by Google Scholar~\cite{google_scholar}, search verified indexed records but mainly rely on keyword matching. Semantic retrieval systems, including OpenScholar~\cite{asai2024openscholar} and Bohrium Science Navigator~\cite{zhang2025bohriumscimaster}, use semantic matching or retrieval-augmented scientific search. Generative LLM baselines include DeepSeek, Kimi, MiniMax, GLM, Gemini and GPT~\cite{deepseekai2025deepseekv32,kimiteam2026kimik2,minimax2025minimaxm1,glm5team2025glm45,google2026gemini31pro,openai2026gpt54}. These LLMs are evaluated as tool-assisted search agents rather than closed-book models: each is equipped with Search and Visit tools and prompted to search, inspect evidence and verify sources before returning papers, following tool-use agent evaluation protocols~\cite{yao2023react,du2026openseeker,du2026openseekerv2}. For fixed-pipeline agentic retrieval, we use Google Scholar Labs~\cite{google2025scholarlabs} as a representative predefined agentic search workflow.

Retrieval quality is measured at $K=20$ using standard information retrieval metrics~\cite{manning2008introduction,jarvelin2002cumulated}. Recall@20 measures how many ground-truth target papers are recovered in the top 20 results; Precision@20 measures how many returned top-20 papers are genuine target papers; F1-score@20 summarizes the balance between recall and precision; and NDCG@20 measures whether relevant papers are ranked closer to the top. We also report per-query cost and source hallucination rate. Cost captures the computational expense of interpreting complex constraints and carrying out retrieval, while hallucination rate measures the proportion of returned papers that cannot be matched to a real scholarly record or contain provably wrong source metadata. Together, these metrics assess intent comprehension, source authenticity and cost-effective retrieval.

\subsection{Accurate, hallucination-free retrieval at low cost}

\begin{table*}[!tbp]
\centering
\caption{Performance comparison of PaSaMaster against other methods. With zero hallucinations guaranteed, PaSaMaster achieves strong recall and ranking with low cost. The generative LLM baselines are allowed to call Search and Visit tools to find papers on the web before returning their final paper lists.}
\label{tab:main_results}
\setlength{\tabcolsep}{4pt}
\renewcommand{\arraystretch}{1.1}
\definecolor{catgray}{gray}{0.92}
\scalebox{0.8}{
\begin{tabular}{lcccccc}
\toprule
\textbf{Method} & \textbf{NDCG} & \textbf{Recall} & \textbf{Precision} & \textbf{F1-score} & \textbf{Hallucination} & \textbf{Cost (\$)} \\
\midrule

\rowcolor{catgray}
\multicolumn{7}{c}{\textit{Lexical Retrieval Systems}} \\
Google Scholar            & 2.07  & 1.69  & 1.48  & 1.39  & 0     & -- \\
\midrule

\rowcolor{catgray}
\multicolumn{7}{c}{\textit{Semantic Retrieval Systems}} \\
OpenScholar               & 14.61 & 11.68 & 8.52  & 7.92  & 0     & -- \\
Bohrium Science Navigator & 22.39 & 19.37 & 12.50 & 12.26 & 0     & -- \\
\midrule

\rowcolor{catgray}
\multicolumn{7}{c}{\textit{Generative LLMs}} \\
DeepSeek-v3.2    & 35.82 & 24.76 & 15.35 & 15.56 & 12.94 & 0.28 \\
Kimi-K2.5        & 37.80 & 28.08 & 16.95 & 17.36 & 26.59 & 0.16 \\
MiniMax-M2.7     & 30.70 & 24.23 & 14.42 & 15.11 & 32.66 & 0.18 \\
GLM-5            & 35.89 & 28.99 & 16.93 & 18.18 & 21.64 & 0.56 \\
Gemini-3.1-pro   & 31.34 & 21.30 & 11.68 & 12.48 & 27.54 & 0.38 \\
GPT-5.2          & 31.59 & 25.32 & 16.82 & 16.69 & 5.65 & 6.06 \\
\midrule

\rowcolor{catgray}
\multicolumn{7}{c}{\textit{Fixed-Pipeline Agentic
Retrieval}} \\
Google Scholar Labs  & 30.54 & 29.01 & 18.79 & 18.87 & 0     & -- \\

\midrule
\rowcolor{catgray}
\multicolumn{7}{c}{\textit{Recursive Self-Evolving Agentic
Retrieval}} \\
\textbf{PaSaMaster}  & \textbf{39.52}& \textbf{33.24} & \textbf{23.46} & \textbf{23.00} & \textbf{0}     & \textbf{0.05} \\
\bottomrule
\end{tabular}
}
\end{table*}

\begin{figure*}[!tbp]
    \centering 
    \begin{minipage}{\linewidth}
        \textbf{a}\par\vspace{0.2em}
        \includegraphics[width=1.0\linewidth]{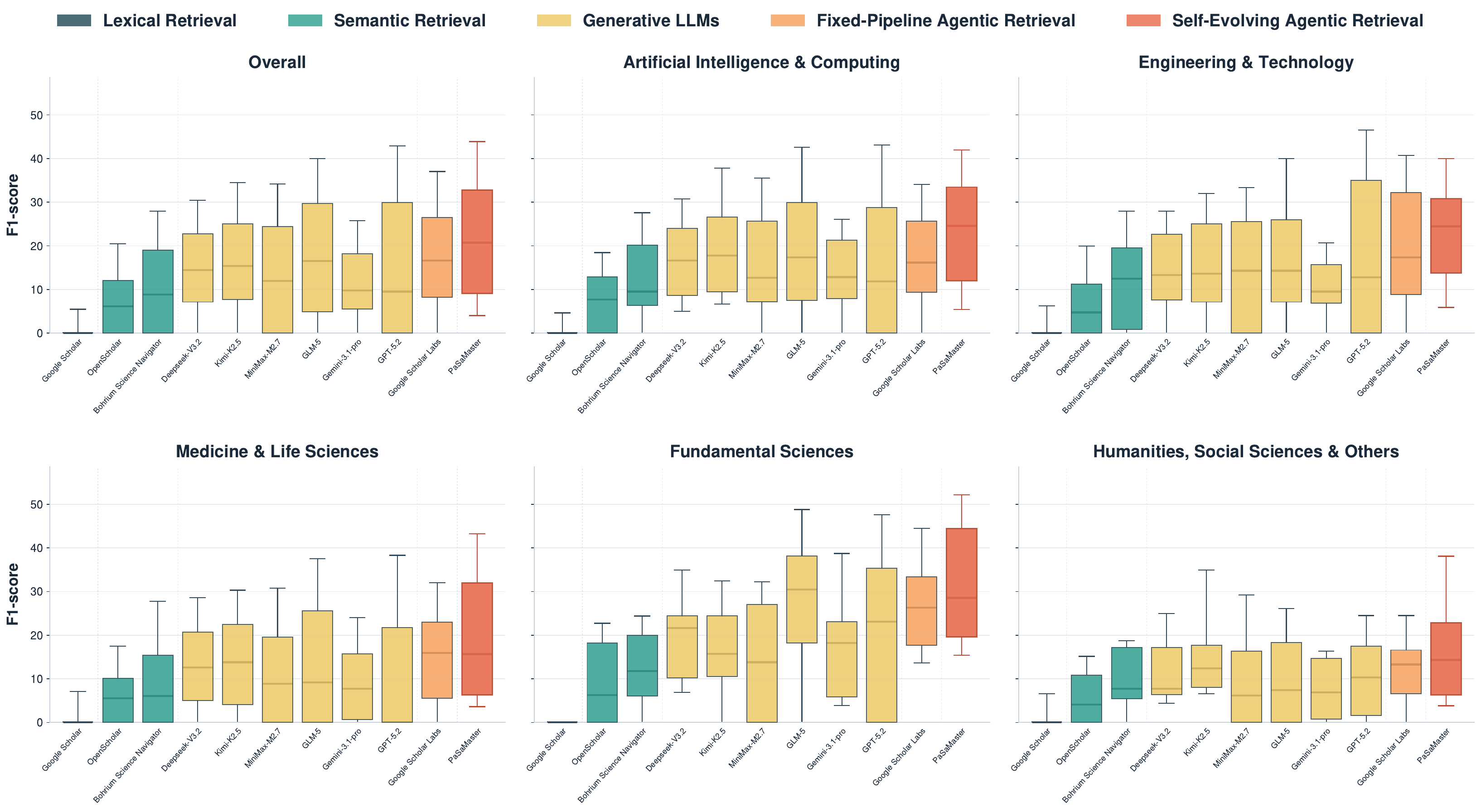}
    \end{minipage}
    \vspace{0.2em}
    \begin{minipage}{\linewidth}
        \textbf{b}\par\vspace{0.2em}
        \includegraphics[width=1.0\linewidth]{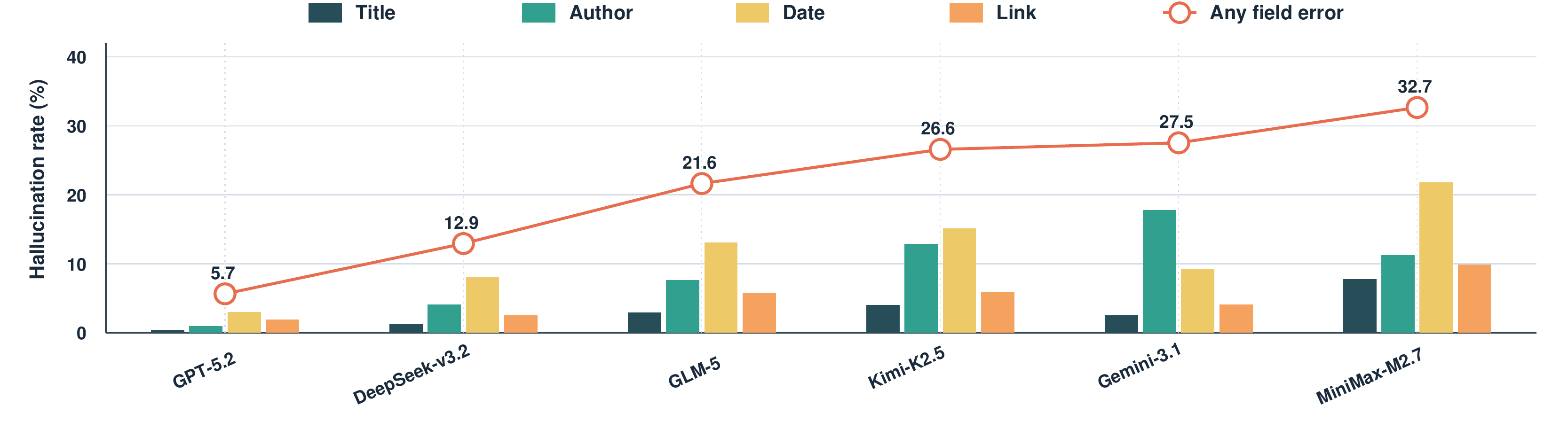}
    \end{minipage}
    \vspace{0.2em}
    \begin{minipage}{\linewidth}
        \textbf{c}\par\vspace{0.2em}
        \includegraphics[width=1.0\linewidth]{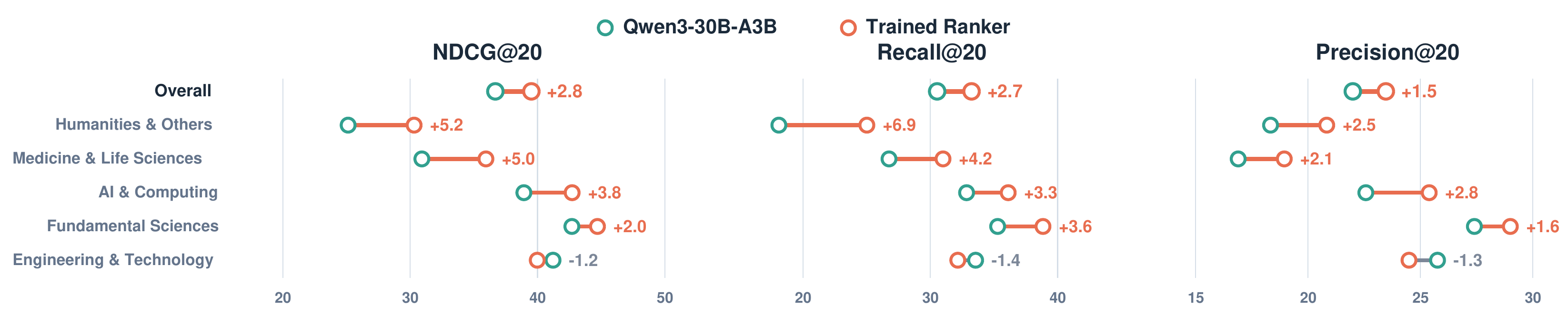}
    \end{minipage}
    \caption{\textbf{Retrieval quality, source hallucination and ranking performance on PaSaMaster-Bench.} \textbf{a,} F1-score distributions comparing different retrieval systems across scientific disciplines. PaSaMaster achieves the strongest overall retrieval performance and leads broadly across subject areas, with both a higher lower bound and a higher upper performance range than competing paradigms. \textbf{b,} Citation-field hallucination rates of tool-assisted generative LLMs, ordered by overall source hallucination rate. Bars show field-specific title, author, date and link errors, and the line marks the overall proportion of returned papers with any source error. \textbf{c,} Overall and per-discipline before--after ranking performance after Ranker training, measured by NDCG@20, Recall@20 and Precision@20. The overall gain is shown first, followed by subject groups ordered by average gain, highlighting broad improvements after multidisciplinary Ranker training.}
    \label{fig:result}
\end{figure*}

Table~\ref{tab:main_results} reports the main results on PaSaMaster-Bench, and Fig.~\ref{fig:result} summarizes cross-disciplinary robustness, source-error patterns and Ranker gains. Overall, PaSaMaster achieves the best retrieval quality while maintaining zero source hallucination and low computational cost. These results support the three central claims of our system: Recursive Self-Evolving retrieval improves understanding of complex natural-language search intents, intent--paper relevance ranking prevents hallucinated sources, and planning--retrieval separation enables cost-efficient large scale literature discovery.

\begin{figure*}[!tbp]
    \centering
    \begin{minipage}{\linewidth}
        \textbf{a}\par\vspace{0.2em}
        \includegraphics[width=1.0\linewidth]{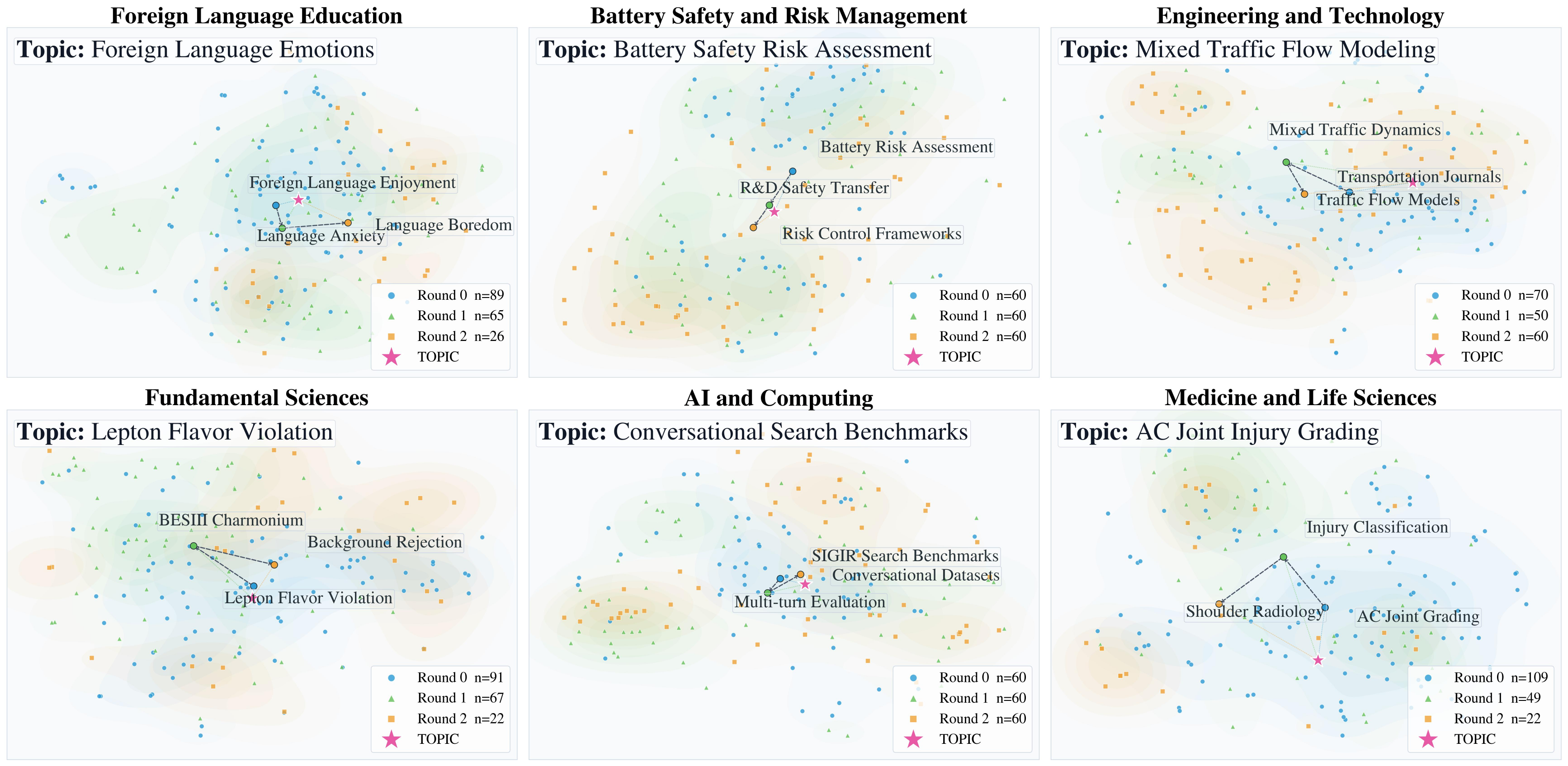}
    \end{minipage}
    \vspace{0.2em}
    \begin{minipage}{\linewidth}
        \textbf{b}\par\vspace{0.05em}
        \centering
        \includegraphics[width=1\linewidth]{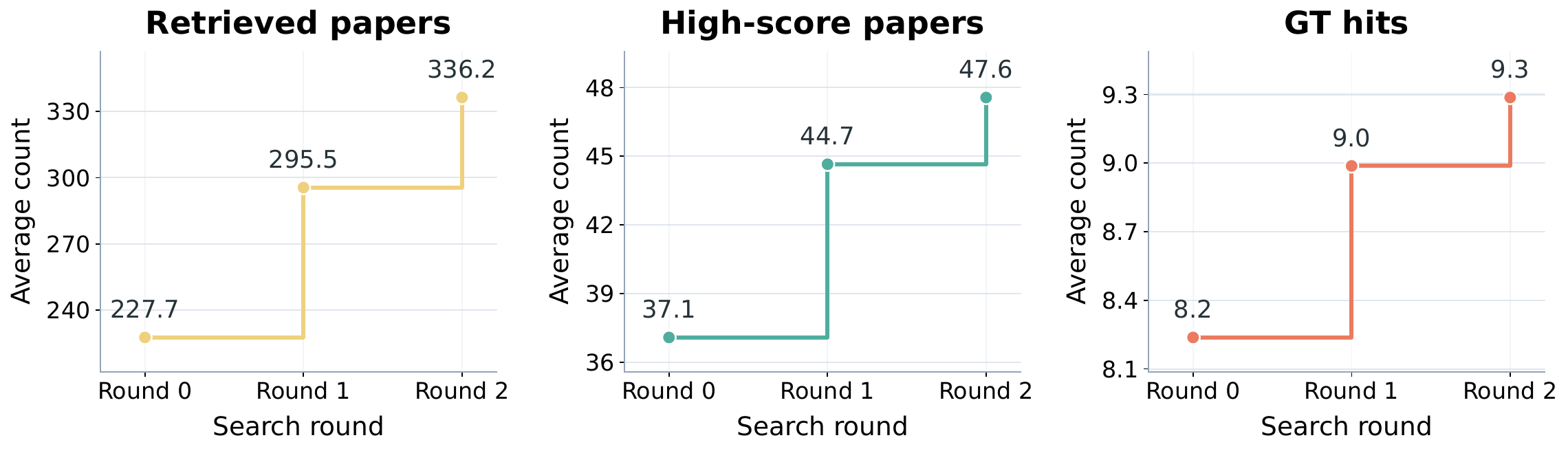}
    \end{minipage}
    \caption{\textbf{Evidence feedback drives intent evolution and target-paper discovery.}
\textbf{a,} t-SNE visualizations of six representative retrieval cases across successive search rounds. Each color denotes papers retrieved in one round, the star marks the original query, density shading indicates retrieved-paper concentration and annotations identify dominant topics. Across rounds, retrieved papers shift toward different topic regions and form distinct distributions, showing that PaSaMaster develops new understanding of the user intent from previously retrieved evidence and uses it to explore new search directions. This evolving cognition helps later rounds move beyond the initial query interpretation and retrieve complementary relevant papers.
\textbf{b,} Retrieval dynamics averaged over all disciplines. Retrieved papers from each round are fed back to the Navigator to refine its understanding of the user intent and update the next search strategy. As the self-evolving process proceeds, PaSaMaster retrieves a larger candidate pool, identifies more high-score papers and recovers more ground-truth target papers. The increase in ground-truth hits across rounds shows that each iteration contributes additional relevant papers rather than merely repeating earlier retrieval results.}
\label{fig:evolving_cases}
\end{figure*}

\textbf{Recursive Self-Evolving retrieval improves complex intent understanding}.  As shown in Table~\ref{tab:main_results}, PaSaMaster achieves the highest retrieval performance across all main quality metrics, with an NDCG@20 of 39.52, Recall@20 of 33.24, Precision@20 of 23.46, and F1-score@20 of 23.00. This demonstrates that PaSaMaster is better able to recover the target papers implied by complex multi-constraint research intents. Compared with Google Scholar~\cite{google_scholar}, PaSaMaster improves F1-score@20 from 1.39 to 23.00, a 16.5$\times$ improvement, showing the severe limitation of keyword-centric retrieval under complex natural-language queries. Compared with semantic retrieval systems, PaSaMaster also substantially outperforms OpenScholar~\cite{asai2024openscholar} and Bohrium Science Navigator~\cite{zhang2025bohriumscimaster}, indicating that passive semantic matching is still insufficient for queries requiring constraint reasoning and intent refinement. Among generative LLMs, the strongest F1-score baseline is GLM-5~\cite{glm5team2025glm45} with 18.18, while the fixed-pipeline agentic retrieval baseline Google Scholar Labs~\cite{google2025scholarlabs} achieves 18.87. PaSaMaster reaches 23.00, improving over these strongest baselines by 26.5\% and 21.9\%, respectively. Fig.~\ref{fig:evolving_cases} provides a mechanistic view of this advantage. In Fig.~\ref{fig:evolving_cases}a, papers retrieved in successive rounds shift toward different topic regions, indicating that evidence from earlier rounds changes the system's interpretation of the query and opens new search directions. In Fig.~\ref{fig:evolving_cases}b, the number of recovered ground-truth papers increases across rounds, showing that these new search directions lead to additional relevant papers rather than merely repeating the initial retrieval. Together, the two panels show that Recursive Self-Evolving retrieval improves complex intent understanding by using retrieved evidence to update cognition and guide later searches toward complementary parts of the intended paper set.

\textbf{Intent–paper relevance ranking eliminates source hallucination.}
As shown in Table~\ref{tab:main_results}, PaSaMaster achieves 0\% hallucination while maintaining the strongest retrieval quality. This result directly supports our design choice of treating literature discovery as intent--paper relevance ranking rather than generation. In contrast, the tool-assisted generative LLM baselines~\cite{deepseekai2025deepseekv32,kimiteam2026kimik2,minimax2025minimaxm1,glm5team2025glm45,google2026gemini31pro,openai2026gpt54} exhibit substantial hallucination rates, including 32.66\% for MiniMax-M2.7, 27.54\% for Gemini-3.1, 26.59\% for Kimi-K2.5, 21.64\% for GLM-5, 12.94\% for DeepSeek-v3.2, and 5.65\% for GPT-5.2. These results show that even frontier LLMs with search and visit tools~\cite{nakano2021webgpt,schick2023toolformer,qin2023toolllm,yao2023react,du2026openseeker,du2026openseekerv2} remain vulnerable to fabricating or misreporting scientific sources~\cite{zhang2025sirenssong,Farquhar_2024}. Fig.~\ref{fig:result}b further shows that hallucinations arise from multiple citation fields, including title, author, date, and link errors. By contrast, PaSaMaster ranks only papers retrieved from verified corpora and grounds relevance judgments in original paper evidence, thereby ensuring zero hallucination in source information.

\textbf{Planning--retrieval separation reduces cost}. Table~\ref{tab:main_results} shows that PaSaMaster achieves this performance at a cost of only \$0.05 per query. This is far below GPT-5.2~\cite{openai2026gpt54} at \$6.06, GLM-5~\cite{glm5team2025glm45} at \$0.56, Gemini-3.1-pro~\cite{google2026gemini31pro} at \$0.38, and DeepSeek-v3.2~\cite{deepseekai2025deepseekv32} at \$0.28. In particular, PaSaMaster outperforms GPT-5.2 in F1-score@20 by 37.8\% while using only about 1\% of its computational cost. This confirms the benefit of separating high-level planning from large-scale retrieval, enabling PaSaMaster to maintain high-quality retrieval at substantially lower cost.

\textbf{Consistent gains across disciplines}. Fig.~\ref{fig:result} reports retrieval robustness, source hallucination and ranking performance across the 38 scientific disciplines in PaSaMaster-Bench. In Fig.~\ref{fig:result}a, PaSaMaster achieves leading F1-score distributions across multiple subject groups. Its performance range is shifted upward relative to competing paradigms, with a higher retrieval floor on difficult cases and a higher upper range on easier cases, indicating that the system improves both robustness and peak retrieval quality across disciplines. In Fig.~\ref{fig:result}c, the overall row shows consistent gains after Ranker training in NDCG@20, Recall@20 and Precision@20, while the subject rows show improvements across most disciplines. These gains reflect training on tens of thousands of multidisciplinary query--paper examples, which helps the Ranker generalize beyond a single domain and place relevant papers more accurately in the final list. Fig.~\ref{fig:evolving_cases}b further shows that Recursive Self-Evolving retrieval expands the evidence space round by round across disciplines, increasing the average number of retrieved papers, high-score papers, and recovered ground-truth papers. Together, these results suggest that PaSaMaster's gains are not driven by a narrow domain-specific advantage. Instead, they reflect the generality of the three design principles: Recursive Self-Evolving retrieval supports complex intent understanding, evidence-grounded ranking ensures source authenticity, and planning--retrieval separation enables scalable retrieval across heterogeneous scientific domains.

\section{Discussion}\label{sec5}
PaSaMaster shows that scientific literature discovery can be framed as a Recursive Self-Evolving, evidence-grounded ranking problem rather than as either keyword matching or citation generation. Across PaSaMaster-Bench, this formulation improves the recovery of target papers under complex search intents while maintaining zero source hallucination and substantially reducing computational cost. This is important because current systems face a structural trade-off: database-backed retrieval preserves source authenticity but often misses the intent behind a nuanced research need, whereas frontier LLMs can reason over richer requests but may fabricate or misreport scientific sources. PaSaMaster narrows this gap by keeping the output space restricted to verified papers while allowing the search process itself to evolve.

The central mechanism is to use ranked evidence to continually update the system's understanding of the user's intent during retrieval. This makes PaSaMaster Recursive Self-Evolving: the search process does not merely execute an initial query, but progressively revises what the query means as evidence accumulates. Researchers rarely know the exact best query before seeing the literature; they search, inspect partial results, recognize missing constraints and refine their direction. PaSaMaster operationalizes this process by letting the Navigator revise the retrieval strategy and verification checklist after observing scored candidates from the Librarian swarm. The resulting loop helps the system move beyond a static interpretation of the initial query, which is especially valuable for multi-constraint intents in which relevance depends on combinations of topic, method, dataset, application context and exclusion criteria. The cross-disciplinary results suggest that this mechanism is not only a domain-specific optimization, but a general strategy for aligning retrieval with complex scientific needs.

Equally important is the decision to rank papers rather than generate citations. In many LLM-based literature workflows, hallucination arises because the model is asked to produce bibliographic objects directly from parametric memory or from partially grounded context. PaSaMaster changes this failure mode by requiring every candidate to be retrieved from a verified corpus and every relevance judgment to be supported by evidence from the original paper. The system therefore does not promise that every relevance judgment is perfect, but it does make each relevance score traceable to paper-level evidence and ensures that recommended sources are real and auditable. This distinction matters for scientific use because fabricated papers can distort a literature review, create false evidence chains and mislead subsequent hypothesis formation or experiment design. PaSaMaster reduces this risk by making each recommendation traceable to a real paper and to the evidence used to judge its relevance.

The low-cost design also shapes the potential use of PaSaMaster. Literature discovery is not a one-off task; researchers repeatedly search while designing projects, writing related work, checking novelty and updating reviews. A system that relies on frontier LLMs for every retrieval and scoring operation is difficult to deploy at this frequency and scale. By separating high-level planning from large-scale retrieval and lightweight relevance scoring, PaSaMaster makes agentic literature discovery more practical for broad, multidisciplinary use. In this role, the system should be viewed as an assistive layer that expands and organizes the candidate evidence space, not as a replacement for expert judgment. Its value is strongest when it delivers more comprehensive and accurate retrieval at low cost, exposes why papers were recommended and makes omissions easier to diagnose.

Several limitations remain. First, PaSaMaster-Bench is expert-curated, and its target sets may still be influenced by the disciplinary expertise, prior knowledge and search horizon of the annotators. We mitigated this limitation by using multiple retrieval channels to help experts assemble broad candidate pools, and by requiring elementwise checklist scoring so that each candidate paper is verified against the stated intent rather than accepted by impression alone. Second, the current evaluation focuses on top-ranked paper retrieval rather than the quality of downstream literature synthesis, hypothesis generation or manuscript writing. Future work should therefore test whether improved retrieval leads to better scientific outputs in human-in-the-loop settings, including expert assessments of coverage, novelty, usefulness and trustworthiness. Third, PaSaMaster could be extended from retrieving relevant papers to helping scientists reconstruct the historical development of a field, identify emerging research trajectories and generate candidate research directions grounded in the literature. Such capabilities would help researchers rapidly understand how a topic has evolved and use verified evidence to define new research questions. Finally, although PaSaMaster eliminates source hallucination by construction, relevance scoring can still inherit biases from corpora, models and checklist design. A deployment-ready system should therefore expose evidence, uncertainty and failure modes clearly enough for researchers to audit its recommendations. Taken together, these results indicate that self-evolving, verified retrieval is a promising foundation for trustworthy AI-assisted science, but its full value will depend on transparent evaluation, continual corpus updating and careful integration into expert research workflows.

\section{Methods}
\subsection{Overview of PaSaMaster}
\label{sec:method}

PaSaMaster is an agentic Recursive Self-Evolving literature retrieval system that maps a complex natural-language search intent $q$ to a ranked, evidence-grounded paper set $\mathcal{P}=[p_1,p_2,\ldots,p_k]$, where $p_i$ is the $i$th recommended paper and $k$ is the output list length. Its design follows three principles that directly address the limitations of existing literature retrieval paradigms: \textit{Recursive Self-Evolving retrieval}, \textit{hallucination-free intent--paper relevance ranking}, and \textit{cost-efficient planning--retrieval separation}. Rather than generating paper lists from parametric memory, PaSaMaster retrieves real papers from customized scientific corpora, verifies their relevance using original evidence, and iteratively refines the search intent based on ranked retrieval results.

Formally, PaSaMaster operates over a customized scientific corpus $\mathcal{D}$ and an agent-accessible operator toolset $\mathcal{T}$. Given query $q$, a Navigator with policy $\pi_{\text{Nav}}$ first produces a retrieval strategy $S$ and a query-specific verification checklist $C=[c_1,c_2,\ldots,c_m]$, where $m$ is the number of checklist items and each checkpoint $c_j$ encodes one concrete requirement that a relevant paper must satisfy. The system also uses $N$ parallel Librarian agents, denoted by $\{\pi_{\text{Lib}}^{(i)}\}_{i=1}^{N}$, where $\pi_{\text{Lib}}^{(i)}$ is the policy of the $i$th Librarian. Let $\textsc{Plan}$ denote Navigator planning, $\textsc{Retrieve}$ denote corpus search, $\textsc{Verify}$ denote evidence-grounded candidate scoring and $\textsc{Rerank}$ denote final listwise ordering. The overall retrieval pipeline is:
\begin{align}
    \langle S,\, C\rangle        &= \textsc{Plan}(q,\pi_{\text{Nav}}), \label{eq:stage1}\\
    \mathcal{P}_{\text{init}}    &= \textsc{Retrieve}\!\left(S;\mathcal{D},\mathcal{T},\{\pi_{\text{Lib}}^{(i)}\}_{i=1}^{N}\right), \label{eq:stage2}\\
    \mathcal{P}_{\text{scored}}  &= \textsc{Verify}\!\left(\mathcal{P}_{\text{init}},C;\mathcal{D},\mathcal{T},\{\pi_{\text{Lib}}^{(i)}\}_{i=1}^{N}\right), \label{eq:stage3}\\
    \mathcal{P}                  &= \textsc{Rerank}\!\left(\mathcal{P}_{\text{scored}}\right), \label{eq:stage4}
\end{align}
Here $\mathcal{P}_{\text{init}}$ is the initially retrieved candidate set, $\mathcal{P}_{\text{scored}}$ is the evidence-scored candidate set and $\mathcal{P}$ is the final ranked paper list. Equations~\ref{eq:stage1}--\ref{eq:stage4} summarize PaSaMaster as a staged process in which planning defines the search, Librarians retrieve and verify papers from trusted tools, and reranking converts scored candidates into the final recommendation list.

\subsubsection{Recursive Self-Evolving Retrieval from Ranked Evidence}

The first core design of PaSaMaster is to transform literature retrieval from one-shot query--document matching into a Recursive Self-Evolving search process. Existing retrieval systems typically fix their interpretation of the user query at the beginning~\cite{google_scholar,asai2024openscholar,zhang2025bohriumscimaster,google2025scholarlabs,he2025pasa} and execute retrieval under this static understanding. PaSaMaster instead treats retrieval as an iterative process in which ranked evidence is used to update the system's understanding of the research intent.

The process is coordinated by the Navigator agent. Given the initial query $q$, the Navigator first analyzes the user's research intent and generates two outputs: a retrieval strategy $S$, specifying what should be searched, and a verification checklist $C$, specifying how candidate papers should be judged. Let $t$ index the retrieval round, and let $S^{(t)}$, $C^{(t)}$ and $\mathcal{P}_{\text{scored}}^{(t)}$ denote the retrieval strategy, checklist and scored candidate set at round $t$. After each retrieval round, the Navigator inspects the ranked results, identifies missing coverage, ambiguous constraints, or under-explored directions, and refines the strategy and checklist for the next round through the reflection operator $\textsc{Reflect}$:
\begin{equation}
    \langle S^{(t+1)}, C^{(t+1)} \rangle
    =
    \textsc{Reflect}\!\left(q,S^{(t)},C^{(t)},\mathcal{P}_{\text{scored}}^{(t)}\right),
    \label{eq:reflect}
\end{equation}
Equation~\ref{eq:reflect} formalizes the self-evolving step: ranked evidence from round $t$ updates the system's understanding of the query and produces the strategy and checklist used in round $t+1$.

\subsubsection{Hallucination-Free Intent--Paper Relevance Ranking}

The second core design is to prevent hallucinated sources by formulating literature discovery as intent--paper relevance ranking rather than generation. PaSaMaster never asks an LLM to synthesize citations or paper lists directly from parametric memory. Instead, every candidate paper must be retrieved from a verified scientific corpus $\mathcal{D}$, and every relevance judgment must be grounded in traceable evidence from the original paper.

To support verifiable retrieval and evidence grounding, PaSaMaster restructures over $160$ million papers into a three-tier agent-native repository~\cite{lo2020s2orc}. The repository contains $\mathcal{D}^{\text{meta}}$ for structured metadata, $\mathcal{D}^{\text{abs}}$ for abstract-level representations used in coarse semantic filtering and $\mathcal{D}^{\text{chunk}}$ for passage-level evidence chunks segmented from full texts. The corpus is represented as:
\begin{equation}
    \mathcal{D}=\{\mathcal{D}^{\text{meta}},\mathcal{D}^{\text{abs}},\mathcal{D}^{\text{chunk}}\},
    \label{eq:db}
\end{equation}
Equation~\ref{eq:db} defines the corpus layout used to keep metadata retrieval, abstract-level filtering and passage-level evidence grounding separate but jointly accessible to the agents.

For each candidate paper $p$ and checklist item $c_j$, let $\mathcal{D}^{\text{chunk}}_p$ denote the set of evidence chunks belonging to paper $p$, let $e$ denote one candidate chunk in this set, let $\phi(\cdot)$ be the shared text encoder, let $\ell$ denote the number of evidence chunks to retrieve, let $\operatorname*{Top}_{\ell}$ denote selection of the $\ell$ highest-scoring chunks and let $\mathcal{E}_p(c_j)$ denote the top-$\ell$ evidence chunks selected to support the judgment for checklist item $c_j$. The Evidence Chunk Locator retrieves supporting passages by cosine similarity:
\begin{equation}
    \mathcal{E}_p(c_j)=
    \operatorname*{Top}_{\ell,\,e\in\mathcal{D}^{\text{chunk}}_p}
    \cos\!\left(\phi(c_j),\phi(e)\right),
    \label{eq:evidence}
\end{equation}
Equation~\ref{eq:evidence} binds each checklist judgment to explicit textual evidence by selecting the passages in a paper that are most semantically aligned with the requirement being checked.

Each candidate paper is then evaluated by a lightweight Scorer model trained through a multidisciplinary distillation pipeline. We first sample seed topics across disciplines and use them to construct initial literature queries and seed paper collections. The retrieved papers are then clustered to obtain finer-grained topic groups, from which we synthesize realistic search queries and query-specific checklists covering topical, methodological, application metadata constraints and exclusion criteria. Each synthesized query is run through the PaSaMaster retrieval pipeline to obtain candidate papers, producing multidisciplinary query--paper pairs that include positive matches, partial matches and hard negatives. A stronger teacher model then labels each pair according to the checklist, assigning checkpoint-level scores, evidence-grounded rationales and holistic relevance judgments. 

This training process gives the lightweight Scorer broad, expert-style relevance assessment ability across disciplines. At inference time, for every paper $p$ and checkpoint $c_j$, the Scorer outputs a satisfaction score $s_j(p)\in\{1,2,3,4,5\}$ and an evidence-grounded rationale, where larger values indicate stronger satisfaction of the checklist item. For a paper $p$, let $\bar{s}(p)$ denote the average checklist satisfaction score over the $m$ checklist items:
\begin{equation}
    \bar{s}(p)=\frac{1}{m}\sum_{j=1}^{m}s_j(p).
    \label{eq:avg}
\end{equation}
Equation~\ref{eq:avg} summarizes checkpoint-level evidence into a single criterion-level relevance signal for candidate paper $p$.

To incorporate holistic confidence, PaSaMaster also extracts the Scorer model's calibrated output probability $\rho\in[0,1]$ for its overall relevance judgment. Let $\mathcal{S}(p)$ denote the final normalized relevance score for paper $p$. The final relevance score is:
\begin{equation}
    \mathcal{S}(p)=\frac{\bar{s}(p)+\rho}{6},
    \label{eq:score}
\end{equation}
where the denominator 6 normalizes the maximum possible value of $\bar{s}(p)+\rho$, because $\bar{s}(p)\leq5$ and $\rho\leq1$. Equation~\ref{eq:score} combines checklist satisfaction and holistic confidence into one paper-level relevance score. The top candidates are then passed to a listwise reranker for global cross-paper comparison. The final result $\mathcal{P}$ is therefore a relevance-ranked list of real papers, with each recommendation traceable to paper-level evidence.

\subsubsection{Cost-Efficient Planning--Retrieval Separation}

The third core design is planning--retrieval separation, which improves scalability by using frontier LLMs only where they are most valuable. Frontier LLMs are effective for understanding, decomposing, and refining complex research intents, but using them for every retrieval, reading, and ranking operation would be unnecessarily expensive. PaSaMaster therefore assigns high-level reasoning to the Navigator and delegates large-scale retrieval and relevance scoring to customized corpora, and lightweight parallel Librarian agents.

Let $\mathcal{T}^{\text{ret}}$ denote the retrieval tools used to construct candidate pools, and let $\mathcal{T}^{\text{read}}$ denote the reading tools used to inspect metadata, abstracts and evidence chunks. The operator toolset is divided into these two subsets:
\begin{equation}
    \mathcal{T}=\mathcal{T}^{\text{ret}}\cup\mathcal{T}^{\text{read}}.
    \label{eq:tools}
\end{equation}
Equation~\ref{eq:tools} states that PaSaMaster separates tools for finding candidate papers from tools for reading and verifying them, which supports cost-efficient division of labor.

The retrieval tools $\mathcal{T}^{\text{ret}}$ construct a broad candidate pool through complementary retrieval channels, including Semantic Direct Retrieval, Citation Network Expansion, and Web-to-Repository Verification. Let $o$ denote one retrieval operator in $\mathcal{T}^{\text{ret}}$, let $\textsc{Retrieve}_{o}(S,\mathcal{D})$ denote the candidates returned by operator $o$ under strategy $S$ over corpus $\mathcal{D}$, and let $\mathcal{C}_{\text{init}}$ denote the union of candidates returned by all retrieval operators:
\begin{equation}
    \mathcal{C}_{\text{init}}=
    \bigcup_{o\in\mathcal{T}^{\text{ret}}}
    \textsc{Retrieve}_{o}(S,\mathcal{D}).
    \label{eq:init}
\end{equation}
Equation~\ref{eq:init} defines the initial candidate pool as the union of complementary retrieval channels, increasing coverage before evidence verification and reranking. Semantic Direct Retrieval provides high-precision semantic candidates, Citation Network Expansion follows citation links to surface structurally related papers, and Web-to-Repository Verification maps external web findings back to verified repository entries. The reading tools $\mathcal{T}^{\text{read}}$ then support efficient metadata lookup, abstract reading, and evidence-chunk localization, avoiding expensive full-document reading and substantially reducing computational cost.

Finally, the distilled Scorer serves as the verification component of each Librarian agent. Given a query-specific checklist and retrieved evidence chunks, it assigns checklist-level scores, generates evidence-grounded rationales and produces a holistic relevance judgment without repeatedly invoking a frontier LLM for every candidate paper. This enables Librarian agents to reproduce expert-style structured verification at much lower inference cost over large scientific corpora.

\subsection{PaSaMaster-Bench}

PaSaMaster-Bench is designed to evaluate literature retrieval systems under conditions that resemble real scientific paper discovery. In practice, researchers rarely ask only for a topic; they ask natural-language questions that combine scientific scope, methods, application setting, metadata restrictions and exclusions, often while searching across the open web rather than a fixed corpus. A benchmark that omits any of these dimensions can overestimate retrieval ability: short keyword-like queries understate intent reasoning, loose relevance labels do not test full constraint satisfaction, single-domain tasks hide cross-disciplinary fragility, and fixed-corpus settings bypass source verification in real search environments.

Table~\ref{tab:bench_comparison} summarizes this gap. Existing literature-search and research-agent benchmarks~\cite{he2025pasa,kang2025researcharenabenchmarkinglargelanguage,bragg2026astabenchrigorousbenchmarkingai,xiong2026autoresearchbenchbenchmarkingaiagents,ajith2024litsearch} cover useful pieces of the problem, but typically miss at least one property needed for realistic paper discovery: complex compositional natural-language intents, broad multidisciplinary coverage, or real-web search. PaSaMaster-Bench is constructed to satisfy all four properties simultaneously. It contains \textbf{244 independent tasks} across \textbf{38 scientific disciplines}, and each task pairs a realistic natural-language search intent with an expert-annotated target paper set $\mathcal{P}^{*}$.

This design makes PaSaMaster-Bench a direct testbed for the capabilities that PaSaMaster is built to provide. A system must infer the user's complete intent, search in realistic environments, return authentic papers and filter candidates by paper-level evidence. A paper is counted as correct only if it satisfies all expert-defined checklist criteria, rather than merely sharing the same topic. The benchmark therefore evaluates whether a retrieval system can recover the intended paper set behind a real research need, not simply whether it can retrieve plausible or broadly relevant papers.

\begin{table*}[!tbp]
\centering
\small
\setlength{\tabcolsep}{4pt}
\renewcommand{\arraystretch}{2.05}
\newcommand{\tabcheck}{\normalsize$\checkmark$}
\newcommand{\tabcross}{\normalsize$\times$}
\caption{\textbf{Comparison with existing literature search and research-agent benchmarks.}
The table compares our benchmark with representative benchmarks across four properties required for realistic scientific paper discovery, including natural-language queries, complex compositional search needs, multidisciplinary coverage and real-web search.}
\label{tab:bench_comparison}
\resizebox{\textwidth}{!}{
\begin{tabular}{@{}>{\raggedright\arraybackslash\normalsize}m{5.1cm}
>{\centering\arraybackslash}m{2.35cm}
>{\centering\arraybackslash}m{2.35cm}
>{\centering\arraybackslash}m{2.35cm}
>{\centering\arraybackslash}m{2.35cm}
>{\centering\arraybackslash}m{2.35cm}
>{\columncolor{gray!10}\centering\arraybackslash}m{2.35cm}@{}}
\toprule[1.2pt]
\rowcolor{gray!12}
\textbf{Benchmark Dimension}
& \makecell{\textbf{RealScholar}\\\textbf{Query}\\[-0.2ex]\scriptsize\cite{he2025pasa}}
& \makecell{\textbf{Research}\\\textbf{Arena}\\[-0.2ex]\scriptsize\cite{kang2025researcharenabenchmarkinglargelanguage}}
& \makecell{\textbf{AstaBench}\\\textbf{Paper Finder}\\[-0.2ex]\scriptsize\cite{bragg2026astabenchrigorousbenchmarkingai}}
& \makecell{\textbf{AutoResearch}\\\textbf{Bench}\\[-0.2ex]\scriptsize\cite{xiong2026autoresearchbenchbenchmarkingaiagents}}
& \makecell{\textbf{LitSearch}\\[-0.2ex]\scriptsize\cite{ajith2024litsearch}}
& \makecell{\textbf{PaSaMaster-}\\\textbf{Bench}} \\
\midrule[0.9pt]

Natural-language queries
& \tabcheck
& \tabcheck
& \tabcheck
& \tabcheck
& \tabcheck
& \textbf{\tabcheck} \\
\midrule[0.45pt]

Complex search intent
& \tabcross
& \tabcross
& \tabcross
& \tabcheck
& \tabcheck
& \textbf{\tabcheck} \\
\midrule[0.45pt]

Multidisciplinary coverage
& \tabcross
& \tabcheck
& \tabcheck
& \tabcross
& \tabcross
& \textbf{\tabcheck} \\
\midrule[0.45pt]

Search in a real web environment
& \tabcheck
& \tabcross
& \tabcross
& \tabcross
& \tabcross
& \textbf{\tabcheck} \\
\bottomrule[1.2pt]
\end{tabular}
}
\end{table*}

\subsubsection{Data Curation}

The curation pipeline is built to preserve realism while making evaluation objective (Fig.~\ref{fig:teaser}b). First, domain experts write search intents grounded in authentic research scenarios across the 38 disciplines. The intent is kept in natural language so that the task resembles how a scientist would ask for papers, but it is also decomposed into a checklist of objective criteria. These criteria specify the topical scope, required methods, application setting, dataset or benchmark conditions, publication restrictions and exclusion rules that define the target paper set.

Second, each query is executed through multiple retrieval channels to approximate a real open-web search process. We use web-enabled frontier LLMs, PaSaMaster's native search engine and traditional web search to build an intentionally broad candidate pool. The purpose is not to treat any retriever as ground truth, but to expose experts to a diverse set of possible targets, including papers that may be missed by one search channel. Retrieved papers are verified, deduplicated and mapped into a unified candidate set before annotation.

Third, domain experts annotate candidates against the checklist item by item. A paper is admitted into $\mathcal{P}^{*}$ only when it satisfies every required criterion. Papers that are topically related but fail a method requirement, application setting, metadata constraint or exclusion rule are marked as incorrect. This elementwise annotation converts complex natural-language needs into verifiable target sets while preserving the compositional difficulty of the original query.

\subsubsection{Evaluation Protocol}

The evaluation protocol asks whether a system can reconstruct the target paper set implied by a complex user intent. Given a query, the system must autonomously search, verify and return a ranked list $\mathcal{P}_{\text{agent}}=[p_1,p_2,\ldots,p_k]$, where $p_i$ is the $i$th returned paper. The list is compared with $\mathcal{P}^{*}$, the expert-annotated set of papers satisfying all checklist criteria. This setup makes the evaluation stricter than topical relevance: a returned paper is useful only if it satisfies the complete intent.

Coverage and ranking quality are evaluated with standard ranking metrics. At cutoff $K=20$, Recall@K measures whether the system can cover the intended paper set; Precision@K measures whether returned papers satisfy the expert checklist; F1@K captures the balance between coverage and constraint satisfaction; and NDCG@K measures whether the most relevant target papers are ranked near the top~\cite{manning2008introduction,jarvelin2002cumulated,thakur2021beir}. In this benchmark, these metrics jointly test natural-language intent comprehension, paper-level filtering and ranking under compositional constraints.

Beyond retrieval quality, PaSaMaster-Bench evaluates source authenticity and cost efficiency, two requirements for deployable scientific search systems. Source hallucination rate measures the proportion of returned papers that cannot be matched to a real scholarly record or contain provably wrong source metadata. Token usage and per-query cost measure whether complex retrieval can remain low-overhead enough to scale across repeated, broad use in scientific workflows. For generative LLM baselines, we evaluate tool-assisted literature search rather than direct parametric answering: each model is equipped with Search and Visit tools and prompted to perform ReAct-style search, evidence inspection and source verification before producing a final ranked list~\cite{nakano2021webgpt,schick2023toolformer,qin2023toolllm,yao2023react,du2026openseeker,du2026openseekerv2}. This setting tests whether LLM agents can use external tools to discover real, constraint-satisfying papers rather than merely generate plausible citations.

Thus, PaSaMaster-Bench measures four capabilities required for realistic scientific literature discovery: \textit{intent comprehension}, \textit{constraint-satisfying retrieval}, \textit{ranking quality} and \textit{source authenticity}. These are also the capabilities targeted by PaSaMaster's self-evolving retrieval, evidence-grounded ranking and planning--retrieval separation.

\subsubsection{Search Intent Examples}

PaSaMaster-Bench is designed around natural-language intents that mirror how researchers actually search for literature. The queries cover four common research needs in scientific work: finding a known type of study, locating target papers when only the research problem is clear, preparing a literature review with strict metadata constraints such as venue or time period, and avoiding superficially similar papers that do not meet the actual requirements. Together, these scenarios make the benchmark closer to real literature discovery than keyword-style retrieval tasks. The examples below show how PaSaMaster-Bench combines topical scope, methodological requirements, application context, metadata restrictions and exclusion criteria, requiring systems to recover papers that satisfy the complete intent rather than merely share keywords.

\begin{querybox}
\small\textbf{Example query 1:} \textit{I need papers on identifying RNA modifications using mass spectrometry that employ machine learning algorithms for spectrum annotation. The mass spectrometry approach should use tandem MS with fragmentation patterns.}
\end{querybox}

This is a direct intent query. The user knows what they want to find, but the target is still compositional: relevant papers must jointly concern RNA modification identification, mass spectrometry, machine-learning-based spectrum annotation and tandem-MS fragmentation evidence.

\begin{querybox}
\small\textbf{Example query 2:} \textit{I'm preparing to construct a computational prediction system for regioselectivity of transition metal catalyzed cross-coupling on mutihalogen reactants, but do not know which prediciton model/descriptors should I include in our method. Can you recommend papers with methods to predict such regioselectivity? Do a thorough search for these papers, provide as many papers as you can find.}
\end{querybox}

This is a problem-driven query. The user does not know the exact model family, descriptors or keywords to search for, but clearly states a research bottleneck. A capable retrieval agent must infer the relevant methodological space and translate the user's need into search criteria before recommending papers.

\begin{querybox}
\small\textbf{Example query 3:} \textit{Could you help me find empirical studies on Federated Learning for multi-hospital collaborative training of medical imaging AI models, published in Nature Medicine, Nature Communications, or IEEE Transactions on Medical Imaging between 2023 and October 2025?}
\end{querybox}

This is a metadata-constrained query. The system must retrieve papers on the target topic while also enforcing publication type and time range. It therefore tests whether retrieval can combine semantic relevance with strict paper-level metadata constraints.

\begin{querybox}
\small\textbf{Example query 4:} \textit{Can you help check if there is a paper that has designed a gel electrolyte for batteries, where this gel electrolyte is not for lithium-ion batteries but for lithium-sulfur batteries? This gel electrolyte is not for other systems, but specifically for the PDOL system.}
\end{querybox}

This exclusion-sensitive query requires selecting a specific battery system and electrolyte chemistry while rejecting nearby but incorrect targets, such as lithium-ion batteries or non-PDOL gel electrolytes. The task therefore requires judgment over both positive and negative conditions, not just semantic similarity to the topic.

Together, these examples cover much of the search behavior encountered in real scientific work: directly specified retrieval, exploratory recommendation from a research problem, metadata-restricted paper finding and fine-grained exclusion of false positives. PaSaMaster-Bench therefore evaluates both the realism and the compositional difficulty of literature discovery, requiring systems to understand what the user means, search vertically within a domain and verify whether each candidate paper truly satisfies the intended requirements.

\bibliography{sn-bibliography}

@misc{google_scholar,
  title        = {{Google Scholar}},
  author       = {{Google}},
  year         = {2026},
  howpublished = {\url{https://scholar.google.com/}}
}

@book{manning2008introduction,
  title     = {{Introduction to Information Retrieval}},
  author    = {Manning, Christopher D. and Raghavan, Prabhakar and Sch{\"u}tze, Hinrich},
  publisher = {Cambridge University Press},
  address   = {Cambridge},
  year      = {2008},
  doi       = {10.1017/CBO9780511809071},
  isbn      = {9780521865715}
}

@article{jarvelin2002cumulated,
author = {J\"{a}rvelin, Kalervo and Kek\"{a}l\"{a}inen, Jaana},
title = {Cumulated gain-based evaluation of IR techniques},
year = {2002},
issue_date = {October 2002},
publisher = {Association for Computing Machinery},
address = {New York, NY, USA},
volume = {20},
number = {4},
issn = {1046-8188},
url = {https://doi.org/10.1145/582415.582418},
doi = {10.1145/582415.582418},
journal = {ACM Trans. Inf. Syst.},
month = oct,
pages = {422–446},
numpages = {25}
}

@misc{yao2023react,
      title={ReAct: Synergizing Reasoning and Acting in Language Models}, 
      author={Shunyu Yao and Jeffrey Zhao and Dian Yu and Nan Du and Izhak Shafran and Karthik Narasimhan and Yuan Cao},
      year={2023},
      eprint={2210.03629},
      archivePrefix={arXiv},
      primaryClass={cs.CL},
      url={https://arxiv.org/abs/2210.03629}, 
}

@misc{du2026openseeker,
      title={OpenSeeker: Democratizing Frontier Search Agents by Fully Open-Sourcing Training Data}, 
      author={Yuwen Du and Rui Ye and Shuo Tang and Xinyu Zhu and Yijun Lu and Yuzhu Cai and Siheng Chen},
      year={2026},
      eprint={2603.15594},
      archivePrefix={arXiv},
      primaryClass={cs.AI},
      url={https://arxiv.org/abs/2603.15594}, 
}

@article{fortunato2018science,
  title   = {Science of science},
  author  = {Fortunato, Santo and Bergstrom, Carl T. and B{\"o}rner, Katy and Evans, James A. and Helbing, Dirk and Milojevi{\'c}, Sta{\v{s}}a and Petersen, Alexander M. and Radicchi, Filippo and Sinatra, Roberta and Uzzi, Brian and Vespignani, Alessandro and Waltman, Ludo and Wang, Dashun and Barab{\'a}si, Albert-L{\'a}szl{\'o}},
  journal = {Science},
  volume  = {359},
  number  = {6379},
  pages   = {eaao0185},
  year    = {2018},
  month   = mar,
  doi     = {10.1126/science.aao0185}
}

@article{bornmann2015growth,
      title={Growth rates of modern science: A bibliometric analysis based on the number of publications and cited references}, 
      author={Lutz Bornmann and Ruediger Mutz},
      year={2014},
      eprint={1402.4578},
      archivePrefix={arXiv},
      primaryClass={cs.DL},
      url={https://arxiv.org/abs/1402.4578}, 
}

@article{Landhuis_2016,
  title   = {Scientific literature: Information overload},
  author  = {Landhuis, Esther},
  journal = {Nature},
  volume  = {535},
  number  = {7612},
  pages   = {457--458},
  year    = {2016},
  url     = {https://www.nature.com/articles/535457a}
}

@misc{houssard2025gerontocratization,
      title={The Gerontocratization of Science: How Hypergrowth Reshapes Knowledge Circulation},
      author={Houssard, Antoine and Gargiulo, Floriana and Venturini, Tommaso and Tubaro, Paola and Di Bona, Gabriele},
      year={2025},
      eprint={2410.00788},
      archivePrefix={arXiv},
      primaryClass={cs.DL},
      url={https://arxiv.org/abs/2410.00788}
}

@book{white2009exploratory,
  title     = {Exploratory Search: Beyond the Query-Response Paradigm},
  author    = {White, Ryen W. and Roth, Resa A.},
  publisher = {Morgan \& Claypool Publishers},
    address   = {San Rafael, CA},
  year      = {2009},
  doi       = {10.2200/S00174ED1V01Y200901ICR003}
}

@article{furnas1987vocabulary,
  title   = {The Vocabulary Problem in Human-System Communication},
  author  = {Furnas, George W. and Landauer, Thomas K. and Gomez, Louis M. and Dumais, Susan T.},
  journal = {Communications of the ACM},
  volume  = {30},
  number  = {11},
  pages   = {964--971},
  year    = {1987},
  doi     = {10.1145/32206.32212}
}

@article{beel2010aseo,
  title   = {Academic Search Engine Optimization ({ASEO}): Optimizing Scholarly Literature for {Google Scholar} and Co.},
  author  = {Beel, J{\"o}ran and Gipp, Bela and Wilde, Erik},
  journal = {Journal of Scholarly Publishing},
  volume  = {41},
  number  = {2},
  pages   = {176--190},
  year    = {2010},
  doi     = {10.3138/jsp.41.2.176}
}

@article{wang2023scientific,
  title   = {Scientific discovery in the age of artificial intelligence},
  author  = {Wang, Hanchen and Fu, T. and Du, Y. and Gao, W. and Huang, K. and Liu, Z. and Chandak, P. and Liu, S. and Van Katwyk, P. and Deac, A. and Anandkumar, A. and Bergen, K. and Gomes, C. P. and Ho, S. and Kohli, P. and Lasenby, J. and Leskovec, J. and Liu, T.-Y. and Manrai, A. and Marks, D. and Ramsundar, B. and Song, L. and Sun, J. and Tang, J. and Veli{\v{c}}kovi{\'c}, P. and Welling, M. and Zhang, L. and Coley, C. W. and Bengio, Y. and Zitnik, M.},
  journal = {Nature},
  volume  = {620},
  pages   = {47--60},
  year    = {2023},
  doi     = {10.1038/s41586-023-06221-2}
}

@misc{he2025pasa,
      title={PaSa: An LLM Agent for Comprehensive Academic Paper Search}, 
      author={Yichen He and Guanhua Huang and Peiyuan Feng and Yuan Lin and Yuchen Zhang and Hang Li and Weinan E},
      year={2025},
      eprint={2501.10120},
      archivePrefix={arXiv},
      primaryClass={cs.IR},
      url={https://arxiv.org/abs/2501.10120}
}

@misc{lewis2021rag,
      title={Retrieval-Augmented Generation for Knowledge-Intensive {NLP} Tasks},
      author={Lewis, Patrick and Perez, Ethan and Piktus, Aleksandra and Petroni, Fabio and Karpukhin, Vladimir and Goyal, Naman and K{\"u}ttler, Heinrich and Lewis, Mike and Yih, Wen-tau and Rockt{\"a}schel, Tim and Riedel, Sebastian and Kiela, Douwe},
      year={2020},
      eprint={2005.11401},
      archivePrefix={arXiv},
      primaryClass={cs.CL},
      url={https://arxiv.org/abs/2005.11401}
}

@misc{karpukhin2020dense,
      title={Dense Passage Retrieval for Open-Domain Question Answering},
      author={Karpukhin, Vladimir and Oguz, Barlas and Min, Sewon and Lewis, Patrick and Wu, Ledell and Edunov, Sergey and Chen, Danqi and Yih, Wen-tau},
      year={2020},
      eprint={2004.04906},
      archivePrefix={arXiv},
      primaryClass={cs.CL},
      url={https://arxiv.org/abs/2004.04906}
}

@misc{zhang2025sirenssong,
      title={Siren's Song in the {AI} Ocean: A Survey on Hallucination in Large Language Models},
      author={Zhang, Yue and Li, Yafu and Cui, Leyang and Cai, Deng and Liu, Lemao and Fu, Tingchen and Huang, Xinting and Zhao, Enbo and Zhang, Yu and Xu, Chen and Chen, Yulong and Wang, Longyue and Luu, Anh Tuan and Bi, Wei and Shi, Freda and Shi, Shuming},
      year={2025},
      eprint={2309.01219},
      archivePrefix={arXiv},
      primaryClass={cs.CL},
      url={https://arxiv.org/abs/2309.01219}
}

@article{Farquhar_2024,
  title   = {Detecting hallucinations in large language models using semantic entropy},
  author  = {Farquhar, Sebastian and Kossen, Jannik and Kuhn, Lorenz and Gal, Yarin},
  journal = {Nature},
  volume  = {630},
  number  = {8017},
  pages   = {625--630},
  year    = {2024},
  doi     = {10.1038/s41586-024-07421-0}
}

@misc{guo2024multiagent,
      title={Large Language Model Based Multi-Agents: A Survey of Progress and Challenges},
      author={Guo, Taicheng and Chen, Xiuying and Wang, Yaqi and Chang, Ruidi and Pei, Shichao and Chawla, Nitesh V. and Wiest, Olaf and Zhang, Xiangliang},
      year={2024},
      eprint={2402.01680},
      archivePrefix={arXiv},
      primaryClass={cs.CL},
      url={https://arxiv.org/abs/2402.01680}
}

@misc{lo2020s2orc,
      title={{S2ORC}: The Semantic Scholar Open Research Corpus},
      author={Lo, Kyle and Wang, Lucy Lu and Neumann, Mark and Kinney, Rodney and Weld, Dan S.},
      year={2020},
      eprint={1911.02782},
      archivePrefix={arXiv},
      primaryClass={cs.CL},
      url={https://arxiv.org/abs/1911.02782}
}

@misc{schick2023toolformer,
      title={Toolformer: Language Models Can Teach Themselves to Use Tools},
      author={Schick, Timo and Dwivedi-Yu, Jane and Dess{\`i}, Roberto and Raileanu, Roberta and Lomeli, Maria and Hambro, Eric and Zettlemoyer, Luke and Cancedda, Nicola and Scialom, Thomas},
      year={2023},
      eprint={2302.04761},
      archivePrefix={arXiv},
      primaryClass={cs.CL},
      url={https://arxiv.org/abs/2302.04761}
}

@misc{qin2023toolllm,
      title={{ToolLLM}: Facilitating Large Language Models to Master 16000+ Real-world {APIs}},
      author={Qin, Yujia and Liang, Shihao and Ye, Yining and Zhu, Kunlun and Yan, Lan and Lu, Yaxi and Lin, Yankai and Cong, Xin and Tang, Xiangru and Qian, Bill and Zhao, Sihan and Tian, Runchu and Xie, Ruobing and Zhou, Jie and Gerstein, Mark and Li, Dahai and Liu, Zhiyuan and Sun, Maosong},
      year={2023},
      eprint={2307.16789},
      archivePrefix={arXiv},
      primaryClass={cs.AI},
      url={https://arxiv.org/abs/2307.16789}
}

@misc{shinn2023reflexion,
      title={Reflexion: Language Agents with Verbal Reinforcement Learning},
      author={Shinn, Noah and Cassano, Federico and Gopinath, Ashwin and Narasimhan, Karthik and Yao, Shunyu},
      year={2023},
      eprint={2303.11366},
      archivePrefix={arXiv},
      primaryClass={cs.AI},
      url={https://arxiv.org/abs/2303.11366}
}

@misc{nakano2021webgpt,
      title={{WebGPT}: Browser-assisted Question-Answering with Human Feedback},
      author={Nakano, Reiichiro and Hilton, Jacob and Balaji, Suchir and Wu, Jeff and Ouyang, Long and Kim, Christina and Hesse, Christopher and Jain, Shantanu and Kosaraju, Vineet and Saunders, William and Jiang, Xu and Cobbe, Karl and Eloundou, Tyna and Krueger, Gretchen and Button, Kevin and Knight, Matthew and Chess, Benjamin and Schulman, John},
      year={2021},
      eprint={2112.09332},
      archivePrefix={arXiv},
      primaryClass={cs.CL},
      url={https://arxiv.org/abs/2112.09332}
}

@misc{thakur2021beir,
      title={{BEIR}: A Heterogeneous Benchmark for Zero-shot Evaluation of Information Retrieval Models},
      author={Thakur, Nandan and Reimers, Nils and R{\"u}ckl{\'e}, Andreas and Srivastava, Abhishek and Gurevych, Iryna},
      year={2021},
      eprint={2104.08663},
      archivePrefix={arXiv},
      primaryClass={cs.IR},
      url={https://arxiv.org/abs/2104.08663}
}

@misc{openai2026gpt54,
  title={Introducing {GPT-5.4}}, 
  author={OpenAI},
  year={2026},
  url={https://openai.com/zh-Hans-CN/index/introducing-gpt-5-4/}
}

@misc{google2026gemini31pro,
  title={{Gemini 3.1 Pro}: A smarter model for your most complex tasks}, 
  author={{Google DeepMind}},
  year={2026},
  url={https://blog.google/innovation-and-ai/models-and-research/gemini-models/gemini-3-1-pro/}
}

@article{du2026openseekerv2,
  title={OpenSeeker-v2: Pushing the Limits of Search Agents with Informative and High-Difficulty Trajectories},
  author={Du, Yuwen and Ye, Rui and Tang, Shuo and Huang, Keduan and Zhu, Xinyu and Cai, Yuzhu and Chen, Siheng},
  journal={arXiv preprint arXiv:2605.04036},
  year={2026}
}

@misc{asai2024openscholar,
    title={OpenScholar: Synthesizing Scientific Literature with Retrieval-augmented LMs}, 
    author={Akari Asai and Jacqueline He and Rulin Shao and Weijia Shi and Amanpreet Singh and Joseph Chee Chang and Kyle Lo and Luca Soldaini and Sergey Feldman and Mike D'arcy and David Wadden and Matt Latzke and Minyang Tian and Pan Ji and Shengyan Liu and Hao Tong and Bohao Wu and Yanyu Xiong and Luke Zettlemoyer and Graham Neubig and Dan Weld and Doug Downey and Wen-tau Yih and Pang Wei Koh and Hannaneh Hajishirzi},
    year={2024},
    eprint={2411.14199},
    archivePrefix={arXiv},
    primaryClass={cs.CL},
    url={http://arxiv.org/abs/2411.14199}
}

@misc{zhang2025bohriumscimaster,
    title={Bohrium + SciMaster: Building the Infrastructure and Ecosystem for Agentic Science at Scale}, 
    author={Linfeng Zhang and Siheng Chen and Yuzhu Cai and Jingyi Chai and Junhan Chang and Kun Chen and Zhi X. Chen and Zhaohan Ding and Yuwen Du and Yuanpeng Gao and Yuan Gao and Jing Gao and Zhifeng Gao and Qiangqiang Gu and Yanhui Hong and Yuan Huang and Xi Fang and Xiaohong Ji and Guolin Ke and Zixing Lei and Xinyu Li and Yongge Li and Ruoxue Liao and Hang Lin and Xiaolu Lin and Yuxiang Liu and Xinzijian Liu and Zexi Liu and Jintan Lu and Tingjia Miao and Haohui Que and Weijie Sun and Yanfeng Wang and Bingyang Wu and Tianju Xue and Rui Ye and Jinzhe Zeng and Duo Zhang and Jiahui Zhang and Linfeng Zhang and Tianhan Zhang and Wenchang Zhang and Yuzhi Zhang and Zezhong Zhang and Hang Zheng and Hui Zhou and Tong Zhu and Xinyu Zhu and Qingguo Zhou and Weinan E},
    year={2025},
    eprint={2512.20469},
    archivePrefix={arXiv},
    primaryClass={cs.AI},
    url={https://arxiv.org/abs/2512.20469}
}

@misc{deepseekai2025deepseekv32,
    title={DeepSeek-V3.2: Pushing the Frontier of Open Large Language Models}, 
    author={DeepSeek-AI and Aixin Liu and Aoxue Mei and Bangcai Lin and Bing Xue and Bingxuan Wang and Bingzheng Xu and Bochao Wu and Bowei Zhang and Chaofan Lin and Chen Dong and Chengda Lu and Chenggang Zhao and Chengqi Deng and Chenhao Xu and Chong Ruan and Damai Dai and Daya Guo and Dejian Yang and Deli Chen and Erhang Li and Fangqi Zhou and Fangyun Lin and Fucong Dai and Guangbo Hao and Guanting Chen and Guowei Li and H. Zhang and Hanwei Xu and Hao Li and Haofen Liang and Haoran Wei and Haowei Zhang and Haowen Luo and Haozhe Ji and Honghui Ding and Hongxuan Tang and Huanqi Cao and Huazuo Gao and Hui Qu and Hui Zeng and Jialiang Huang and Jiashi Li and Jiaxin Xu and Jiewen Hu and Jingchang Chen and Jingting Xiang and Jingyang Yuan and Jingyuan Cheng and Jinhua Zhu and Jun Ran and Junguang Jiang and Junjie Qiu and Junlong Li and Junxiao Song and Kai Dong and Kaige Gao and Kang Guan and Kexin Huang and Kexing Zhou and Kezhao Huang and Kuai Yu and Lean Wang and Lecong Zhang and Lei Wang and Liang Zhao and Liangsheng Yin and Lihua Guo and Lingxiao Luo and Linwang Ma and Litong Wang and Liyue Zhang and Mingchuan Zhang and Minghua Zhang and Minghui Tang and Mingxu Zhou and Panpan Huang and Peixin Cong and Peiyi Wang and Qiancheng Wang and Qihao Zhu and Qingyang Li and Qinyu Chen and Qiushi Du and Ruiling Xu and Ruiqi Ge and Ruisong Zhang and Ruizhe Pan and Runji Wang and Runqiu Yin and Runxin Xu and Ruomeng Shen and Ruoyu Zhang and Shanghao Lu and Shangyan Zhou and Shanhuang Chen and Shaofei Cai and Shaoyuan Chen and Shengding Hu and Shengyu Liu and Shiqiang Hu and Shirong Ma and Shiyu Wang and Shuiping Yu and Shunfeng Zhou and Shuting Pan and Songyang Zhou and Tao Ni and Tao Yun and Tian Pei and Tian Ye and Tianyuan Yue and Wangding Zeng and Wen Liu and Wenfeng Liang and Wenjun Gao and Wentao Zhang and Xiao Bi and Xiaodong Liu and Xiaohan Wang and Xiaokang Chen and Xiaokang Zhang and Xiaotao Nie and Xin Cheng and Xin Liu and Xin Xie and Xingchao Liu and Xingkai Yu and Xinyu Yang and Xinyuan Li and Xuecheng Su and Xuheng Lin and Yang Zhang and Yanhong Xu and Yao Li and Yao Zhao and Yaofeng Sun and Yaohui Wang and Yi Yu and Yichao Zhang and Yifan Shi and Yiliang Xiong and Ying He and Yishi Piao and Yisong Wang and Yixuan Tan and Yiyang Ma and Yiyuan Liu and Yongqiang Guo and Yu Wu and Yuan Ou and Yuduan Wang and Yue Gong and Yuheng Zou and Yunfan Xiong and Yuxiang Luo and Yuxiang You and Yuxuan Liu and Yuyang Zhou and Zehui Ren and Zhangli Sha and Zhe Fu and Zhean Xu and Zhenda Xie and Zhengyan Zhang and Zhewen Hao and Zhibin Gou and Zhicheng Ma and Zhigang Yan and Zhihong Shao and Zhiyu Wu and Zhuoshu Li and Zihui Gu and Zijia Zhu and Zilin Li and Ziwei Xie and Ziyi Gao and Zizheng Pan},
    year={2025},
    eprint={2512.02556},
    archivePrefix={arXiv},
    primaryClass={cs.CL},
    url={https://arxiv.org/abs/2512.02556}
}

@misc{kimiteam2026kimik2,
    title={Kimi K2: Open Agentic Intelligence}, 
    author={Kimi Team and Yifan Bai and Yiping Bao and Y. Charles and Cheng Chen and Guanduo Chen and Haiting Chen and Huarong Chen and Jiahao Chen and Ningxin Chen and Ruijue Chen and Yanru Chen and Yuankun Chen and Yutian Chen and Zhuofu Chen and Jialei Cui and Hao Ding and Mengnan Dong and Angang Du and Chenzhuang Du and Dikang Du and Yulun Du and Yu Fan and Yichen Feng and Kelin Fu and Bofei Gao and Chenxiao Gao and Hongcheng Gao and Peizhong Gao and Tong Gao and Yuyao Ge and Shangyi Geng and Qizheng Gu and Xinran Gu and Longyu Guan and Haiqing Guo and Jianhang Guo and Xiaoru Hao and Tianhong He and Weiran He and Wenyang He and Yunjia He and Chao Hong and Hao Hu and Yangyang Hu and Zhenxing Hu and Weixiao Huang and Zhiqi Huang and Zihao Huang and Tao Jiang and Zhejun Jiang and Xinyi Jin and Yongsheng Kang and Guokun Lai and Cheng Li and Fang Li and Haoyang Li and Ming Li and Wentao Li and Yang Li and Yanhao Li and Yiwei Li and Zhaowei Li and Zheming Li and Hongzhan Lin and Xiaohan Lin and Zongyu Lin and Chengyin Liu and Chenyu Liu and Hongzhang Liu and Jingyuan Liu and Junqi Liu and Liang Liu and Shaowei Liu and Tianwei Liu and Weizhou Liu and Yangyang Liu and Yibo Liu and Yiping Liu and Yue Liu and Zhengying Liu and Enzhe Lu and Haoyu Lu and Lijun Lu and Yashuo Luo and Shengling Ma and Xinyu Ma and Yingwei Ma and Shaoguang Mao and Jie Mei and Xin Men and Yibo Miao and Siyuan Pan and Yebo Peng and Ruoyu Qin and Zeyu Qin and Bowen Qu and Zeyu Shang and Lidong Shi and Shengyuan Shi and Feifan Song and Jianlin Su and Zhengyuan Su and Lin Sui and Xinjie Sun and Flood Sung and Yunpeng Tai and Heyi Tang and Jiawen Tao and Qifeng Teng and Chaoran Tian and Chensi Wang and Dinglu Wang and Feng Wang and Hailong Wang and Haiming Wang and Jianzhou Wang and Jiaxing Wang and Jinhong Wang and Shengjie Wang and Shuyi Wang and Si Wang and Xinyuan Wang and Yao Wang and Yejie Wang and Yiqin Wang and Yuxin Wang and Yuzhi Wang and Zhaoji Wang and Zhengtao Wang and Zhexu Wang and Chu Wei and Qianqian Wei and Haoning Wu and Wenhao Wu and Xingzhe Wu and Yuxin Wu and Chenjun Xiao and Jin Xie and Xiaotong Xie and Weimin Xiong and Boyu Xu and Jinjing Xu and Lin Xu and Suting Xu and Weixin Xu and Xinran Xu and Yangchuan Xu and Ziyao Xu and Jing Xu and Junjie Yan and Yuzi Yan and Hao Yang and Xiaofei Yang and Yi Yang and Ying Yang and Zhen Yang and Zhilin Yang and Zonghan Yang and Haotian Yao and Xingcheng Yao and Wenjie Ye and Zhuorui Ye and Bohong Yin and Longhui Yu and Enming Yuan and Hongbang Yuan and Mengjie Yuan and Siyu Yuan and Haobing Zhan and Dehao Zhang and Hao Zhang and Wanlu Zhang and Xiaobin Zhang and Yadong Zhang and Yangkun Zhang and Yichi Zhang and Yizhi Zhang and Yongting Zhang and Yu Zhang and Yutao Zhang and Yutong Zhang and Zheng Zhang and Haotian Zhao and Yikai Zhao and Zijia Zhao and Huabin Zheng and Shaojie Zheng and Longguang Zhong and Jianren Zhou and Xinyu Zhou and Zaida Zhou and Jinguo Zhu and Zhen Zhu and Weiyu Zhuang and Xinxing Zu},
    year={2026},
    eprint={2507.20534},
    archivePrefix={arXiv},
    primaryClass={cs.LG},
    url={https://arxiv.org/abs/2507.20534}
}

@misc{minimax2025minimaxm1,
    title={MiniMax-M1: Scaling Test-Time Compute Efficiently with Lightning Attention}, 
    author={MiniMax and Aili Chen and Aonian Li and Bangwei Gong and Binyang Jiang and Bo Fei and Bo Yang and Boji Shan and Changqing Yu and Chao Wang and Cheng Zhu and Chengjun Xiao and Chengyu Du and Chi Zhang and Chu Qiao and Chunhao Zhang and Chunhui Du and Congchao Guo and Da Chen and Deming Ding and Dianjun Sun and Dong Li and Enwei Jiao and Haigang Zhou and Haimo Zhang and Han Ding and Haohai Sun and Haoyu Feng and Huaiguang Cai and Haichao Zhu and Jian Sun and Jiaqi Zhuang and Jiaren Cai and Jiayuan Song and Jin Zhu and Jingyang Li and Jinhao Tian and Jinli Liu and Junhao Xu and Junjie Yan and Junteng Liu and Junxian He and Kaiyi Feng and Ke Yang and Kecheng Xiao and Le Han and Leyang Wang and Lianfei Yu and Liheng Feng and Lin Li and Lin Zheng and Linge Du and Lingyu Yang and Lunbin Zeng and Minghui Yu and Mingliang Tao and Mingyuan Chi and Mozhi Zhang and Mujie Lin and Nan Hu and Nongyu Di and Peng Gao and Pengfei Li and Pengyu Zhao and Qibing Ren and Qidi Xu and Qile Li and Qin Wang and Rong Tian and Ruitao Leng and Shaoxiang Chen and Shaoyu Chen and Shengmin Shi and Shitong Weng and Shuchang Guan and Shuqi Yu and Sichen Li and Songquan Zhu and Tengfei Li and Tianchi Cai and Tianrun Liang and Weiyu Cheng and Weize Kong and Wenkai Li and Xiancai Chen and Xiangjun Song and Xiao Luo and Xiao Su and Xiaobo Li and Xiaodong Han and Xinzhu Hou and Xuan Lu and Xun Zou and Xuyang Shen and Yan Gong and Yan Ma and Yang Wang and Yiqi Shi and Yiran Zhong and Yonghong Duan and Yongxiang Fu and Yongyi Hu and Yu Gao and Yuanxiang Fan and Yufeng Yang and Yuhao Li and Yulin Hu and Yunan Huang and Yunji Li and Yunzhi Xu and Yuxin Mao and Yuxuan Shi and Yuze Wenren and Zehan Li and Zelin Li and Zhanxu Tian and Zhengmao Zhu and Zhenhua Fan and Zhenzhen Wu and Zhichao Xu and Zhihang Yu and Zhiheng Lyu and Zhuo Jiang and Zibo Gao and Zijia Wu and Zijian Song and Zijun Sun},
    year={2025},
    eprint={2506.13585},
    archivePrefix={arXiv},
    primaryClass={cs.CL},
    url={https://arxiv.org/abs/2506.13585}
}

@misc{glm5team2025glm45,
    title={GLM-4.5: Agentic, Reasoning, and Coding ({ARC}) Foundation Models}, 
    author={{GLM} Team and Aohan Zeng and Xin Lv and Qinkai Zheng and Zhenyu Hou and Bin Chen and Chengxing Xie and Cunxiang Wang and Da Yin and Hao Zeng and Jiajie Zhang and Kedong Wang and Lucen Zhong and Mingdao Liu and Rui Lu and Shulin Cao and Xiaohan Zhang and Xuancheng Huang and Yao Wei and Yean Cheng and Yifan An and Yilin Niu and Yuanhao Wen and Yushi Bai and Zhengxiao Du and Zihan Wang and Zilin Zhu},
    year={2025},
    eprint={2508.06471},
    archivePrefix={arXiv},
    primaryClass={cs.CL},
    url={https://arxiv.org/abs/2508.06471}
}

@misc{google2025scholarlabs,
    title={Scholar Labs: An {AI} Powered Scholar Search},
    author={{Google}},
    year={2025},
    howpublished={Google Scholar Blog},
    url={https://scholar.googleblog.com/2025/11/scholar-labs-ai-powered-scholar-search.html}
}

@article{gusenbauer2021searching,
  title   = {What every researcher should know about searching---clarified concepts, search advice, and an agenda to improve finding in academia},
  author  = {Gusenbauer, Michael and Haddaway, Neal R.},
  journal = {Research Synthesis Methods},
  volume  = {12},
  number  = {2},
  pages   = {136--147},
  year    = {2021},
  doi     = {10.1002/jrsm.1457}
}

@misc{ajith2024litsearch,
      title={LitSearch: A Retrieval Benchmark for Scientific Literature Search}, 
      author={Anirudh Ajith and Mengzhou Xia and Alexis Chevalier and Tanya Goyal and Danqi Chen and Tianyu Gao},
      year={2024},
      eprint={2407.18940},
      archivePrefix={arXiv},
      primaryClass={cs.IR},
      url={https://arxiv.org/abs/2407.18940}, 
}

@article{zhang2025llmscimethod,
  title   = {Exploring the role of large language models in the scientific method: from hypothesis to discovery},
  author  = {Zhang, Yanbo and Khan, Sumeer A. and Mahmud, Adnan and Yang, Huck and Lavin, Alexander and Levin, Michael and Frey, Jeremy G. and Dunnmon, Jared and Evans, James and Bundy, Alan and D{\v{z}}eroski, Sa{\v{s}}o and Tegn{\'e}r, Jesper and Zenil, Hector},
  journal = {npj Artificial Intelligence},
  volume  = {1},
  number  = {1},
  pages   = {14},
  year    = {2025},
  doi     = {10.1038/s44387-025-00019-5}
}

@online{pubmed-ncbi-official,
  author       = {{National Center for Biotechnology Information}},
  title        = {PubMed},
  medium       = {Internet},
  address      = {Bethesda (MD)},
  publisher    = {National Library of Medicine (US), National Center for Biotechnology Information},
  year         = {1996},
  urldate      = {2026-05-08},
  url          = {https://pubmed.ncbi.nlm.nih.gov/}
}

@misc{kang2025researcharenabenchmarkinglargelanguage,
      title={ResearchArena: Benchmarking Large Language Models' Ability to Collect and Organize Information as Research Agents}, 
      author={Hao Kang and Chenyan Xiong},
      year={2025},
      eprint={2406.10291},
      archivePrefix={arXiv},
      primaryClass={cs.AI},
      url={https://arxiv.org/abs/2406.10291}
}

@misc{bragg2026astabenchrigorousbenchmarkingai,
      title={AstaBench: Rigorous Benchmarking of AI Agents with a Scientific Research Suite}, 
      author={Jonathan Bragg and Mike D'Arcy and Nishant Balepur and Dan Bareket and Bhavana Dalvi and Sergey Feldman and Dany Haddad and Jena D. Hwang and Peter Jansen and Varsha Kishore and Bodhisattwa Prasad Majumder and Aakanksha Naik and Sigal Rahamimov and Kyle Richardson and Amanpreet Singh and Harshit Surana and Aryeh Tiktinsky and Rosni Vasu and Guy Wiener and Chloe Anastasiades and Stefan Candra and Jason Dunkelberger and Dan Emery and Rob Evans and Malachi Hamada and Regan Huff and Rodney Kinney and Matt Latzke and Jaron Lochner and Ruben Lozano-Aguilera and Cecile Nguyen and Smita Rao and Amber Tanaka and Brooke Vlahos and Peter Clark and Doug Downey and Yoav Goldberg and Ashish Sabharwal and Daniel S. Weld},
      year={2026},
      eprint={2510.21652},
      archivePrefix={arXiv},
      primaryClass={cs.AI},
      url={https://arxiv.org/abs/2510.21652}
}

@misc{xiong2026autoresearchbenchbenchmarkingaiagents,
      title={AutoResearchBench: Benchmarking AI Agents on Complex Scientific Literature Discovery}, 
      author={Lei Xiong and Kun Luo and Ziyi Xia and Wenbo Zhang and Jin-Ge Yao and Zheng Liu and Jingying Shao and Jianlyu Chen and Hongjin Qian and Xi Yang and Qian Yu and Hao Li and Chen Yue and Xiaan Du and Yuyang Wang and Yesheng Liu and Haiyu Xu and Zhicheng Dou},
      year={2026},
      eprint={2604.25256},
      archivePrefix={arXiv},
      primaryClass={cs.AI},
      url={https://arxiv.org/abs/2604.25256}
}

\end{document}